\documentclass[twocolumn,prx,showpacs,amsmath]{revtex4-1}
\usepackage[xdvi]{graphicx}
\usepackage{color}
\usepackage{ulem}
\usepackage{mathrsfs}

\newcommand{\pder}[2]{\frac{\partial #1}{\partial  #2}}
\newcommand{\pderf}[3]{\left(\frac{\partial #1}{\partial  #2}\right)_{#3}}

\DeclareMathOperator*{\argmin}{arg\,min} 
\DeclareMathOperator*{\argmax}{arg\,max} 

\newcommand{\intx}{\int_{0}^{L_x} dx~}
\newcommand{\intL}{\int_{0}^{X} dx~}
\newcommand{\intG}{\int_{X}^{L_x} dx~}

\newcommand{\ep}{\varepsilon}

\newcommand{\kB}{k_\mathrm{B}}

\newcommand{\eq}{\mathrm{eq}}

\newcommand{\subC}{\mathrm{c}}
\newcommand{\subM}{\mathrm{m}}
\newcommand{\subL}{\mathrm{L}}
\newcommand{\subG}{\mathrm{G}}
\newcommand{\subLG}{\mathrm{L/G}}

\newcommand{\subl}{\mathrm{\ell}}
\newcommand{\subr}{\mathrm{r}}
\newcommand{\sublr}{\mathrm{\ell/r}}
\newcommand{\Tc}{T_\mathrm{c}}

\newcommand{\pex}{p}

\newcommand{\bT}{{\tilde T}}
\newcommand{\bmu}{\tilde\mu}
\newcommand{\bbeta}{{\tilde \beta}}
\newcommand{\balpha}{{\tilde \alpha}}

\makeatletter
\@addtoreset{equation}{section}
\makeatother

\begin{document}
\title{Unique extension of the maximum entropy principle to phase coexistence in heat conduction}

\author{Naoko Nakagawa}
\affiliation {Department of Physics,  
              Ibaraki University, Mito 310-8512, Japan}

\author{Shin-ichi Sasa}
\affiliation {
Department of Physics, Kyoto University, Kyoto 606-8502, Japan}

\date{\today}

\begin{abstract}
The maximum entropy principle determines the values of thermodynamic variables in thermally isolated equilibrium systems. This paper extends the principle to a variational principle that applies to liquid--gas coexistence in heat conduction. We show the uniqueness of the extension under the assumption that the variational principle and the fundamental thermodynamic relation are simultaneously extended in the linear response regime with the total energy fixed. Using the extended variational principle, we calculate the thermodynamic quantities in this steady state and find that the temperature of the liquid--gas interface deviates from the equilibrium transition temperature, which should be verified in experiments.
\end{abstract}

\maketitle

\section{Introduction}

The values of extensive unconstrained variables in thermally isolated equilibrium systems maximize the total entropy. This variational principle is called the {\it maximum entropy principle} and plays a fundamental role in equilibrium thermodynamics \cite{Callen}. The principle leads to the definition of temperature and provides the equation of state. Additionally, the fluctuation law of thermodynamic variables is naturally conjectured from the variational principle and turns out to be consistent with the principle of equal weight in statistical mechanics \cite{Einstein}. The main question addressed in this paper is whether the maximum entropy principle extends to out-of-equilibrium systems. 

The first approach to answering the above question relies on nonequilibrium statistical mechanics. In the last two decades, several universal properties of systems out of equilibrium have been understood with the aid of the fluctuation theorem and its several variants \cite{Evans-Cohen-Morriss,Gallavotti,JarzynskiPRL,Kurchan,LS,Maes,Crooks-WR,Hatano-Sasa, Sekimoto-book,SeifertRPP}. 
In particular, the difference in the stationary distribution from nonequilibrium to equilibrium has been studied when the Shannon entropy has been assumed to be the nonequilibrium extension of equilibrium entropy.
Specifically, for general systems with local detailed balance, the representation of the stationary distribution has been known to include the time integration of the entropy production in the relaxation to the equilibrium state \cite{Crooks}. In the linear response regime, the representation takes the form proposed by Zubarev and Mclennan \cite{Zubarev,Mclennan}. Additionally, a representation to higher order beyond the linear response regime has been proposed \cite{KN,KNST}. Although these representations are valid, they are not directly used to analyze thermodynamic properties owing to the difficulty of treating the time integration.
For some particular models, stationary distributions without the time integration have been derived \cite{DLS,Jona-Lasinio}. 
However, these forms are too specific to apply to general out-of-equilibrium systems.
There has been no successful extension of the maximum entropy principle based on nonequilibrium statistical mechanics.

The second approach 
is to construct a phenomenological framework of the extension of thermodynamics. There has been a long history of attempting to construct this framework \cite{Keizer,Eu,Jou,Oono-Paniconi,Sasa-Tasaki,Bertin,Seifert-contact,Dickman,Holyst}. All attempts have adopted the same basic idea of adding a nonequilibrium variable and its conjugate variable to equilibrium thermodynamic variables. 
However, the method adopted for the extension depends on the theory, and indeed there are many different versions. For each specific setup, a nonequilibrium variable is chosen by noting the intensive or extensive nature. The conjugate variable of the nonequilibrium variable is defined from the thermodynamic relation using extended entropy. However, to introduce extended entropy, the theory requires an additional assumption, which is at the heart of the argument for each theory. It has not yet been established which assumption is most reasonable, and the extension of the maximum entropy principle, whether or not it is aimed at, is not a requisite for the construction of the framework for each theory.

\begin{figure}[b]
\begin{center}
\includegraphics[scale=0.45]{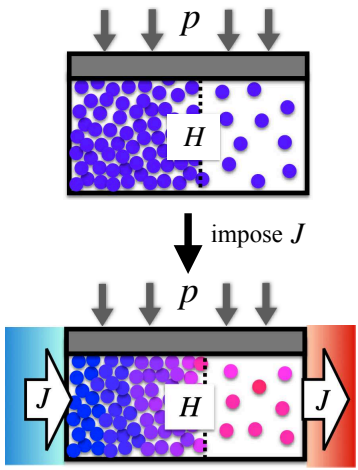}
\caption{Schematic illustration of thermally isolated systems in and out of equilibrium. At constant pressure of $p$, the total conserved energy of the system is enthalpy $H$.  $J$ is the amount of heat current per unit area imposed at the left and right boundary walls. See the main text for further details.}
\label{fig:setup}
\end{center}
\end{figure}

The central concept of the paper is that the maximum entropy principle and the fundamental thermodynamic relation are simultaneously extended. That is, we first assume an extension of the equilibrium variational function with an undetermined function, and then impose the value of the variational function in the steady state to satisfy a fundamental thermodynamic relation.
In cases of equilibrium, the maximum entropy principle is closely related to the law of entropy increase for thermally isolated systems, where the total energy is conserved without external operations. 
 Thus, we should start with the design
of energy-conserving systems in heat conduction.
An  example of such a system is shown in Fig. \ref{fig:setup}, where the in-flow heat current is controlled as it is strictly the same as the out-flow current at any moment \cite{global-fluc}. 
We attempt to formulate an extended maximum entropy principle for determining the thermodynamic quantities of such systems. 

The important thing here is that the extended maximum entropy principle should provide a novel prediction of quantitative phenomena that can be observed in experiments. From this viewpoint, the liquid--gas phase coexistence in heat conduction has turned out to be a good target even in the linear response regime. Our previous papers predicted that the interface temperature deviates from the equilibrium transition temperature \cite{global-PRL,global-JSP}. Numerical simulations have quantitatively confirmed this prediction \cite{KNS}.  
Such phase coexistence may become studied by extending the maximum entropy principle to the linear response regime in energy-conserving heat-conduction systems.

Here, we briefly review the maximum entropy principle for the liquid--gas coexistence at equilibrium.
At constant pressure, 
the thermodynamic entropy $S$ is a function of the energy and the number of particles of the whole system, whereas the variational function $\mathscr S$ is a function of those of the liquid with the total energy and the total number of particles of the system fixed.
The two entropies are functionally different and connected by the maximum entropy principle; i.e.,
the value of $S$ is equal to the maximum value of $\mathscr S$. 
The maximization determines the unique equilibrium state. 
The equilibrium entropy possesses a remarkable property that the two entropies, $\mathscr S$ and $S$, are similarly expressed as the sum of the entropy of the liquid and the entropy of the gas.
 
We proceed to the heat conduction state in the linear response regime.
The respective entropies of the liquid and the gas are given by the integration of the local entropy density in the respective regions.
One may then guess that the variational function $\mathscr S$ is obtained by summing these entropies similarly to the equilibrium cases. However, there is no reason to assume the additivity of entropy for out-of-equilibrium systems. Instead of assuming the additivity, we impose that the value of the variational function in the steady state satisfies the thermodynamic relation with the global temperature introduced in Refs. \cite{global-PRL}.

From this fundamental assumption and other assumptions of the extensive/intensive nature of variables, we can uniquely determine the form of the variational function. The remarkable result obtained in this paper is that the variational function has a non-additive contribution even in the linear response regime. This term is specific to phase coexistence because it is proportional to the latent heat. Furthermore, we show that the steady state determined within this framework is identical to that determined by the free energy minimum principle with the global temperature fixed \cite{global-PRL}. As a consequence, the interface temperature deviates from the equilibrium transition temperature.

The remainder of the paper is organized as follows. Section \ref{s:review} reviews the maximum entropy principle for determining the thermodynamic state of coexisting liquid and gas phases  with the total enthalpy, pressure and total particle number fixed. Section \ref{s:setup} describes the setup for the heat conduction states that we study. Section \ref{s:problem} addresses the problem that we want to solve; i.e., we seek a form of the variational function as an extension of the maximum entropy principle. Section \ref{s:results} presents the form of the variational function, which gives the unique answer to the problem. Section \ref{s:equiv} argues the thermodynamic equivalence of steady states in heat conduction.
Section \ref{s:steady-state} provides a quantitative result for the steady state.  The final section is devoted to concluding remarks. Derivations of the results and formulas are separately presented in the appendices. In this paper, we define the temperature using the energy unit $\kB=1$ for simplicity.

\section{Brief review on the role of entropy in equilibrium}
\label{s:review}

\begin{figure}[bt]
\begin{center}
\includegraphics[scale=0.42]{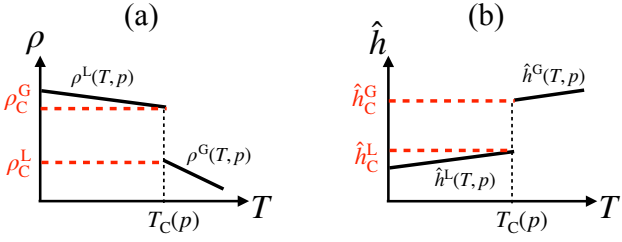}
\caption{
(a) Particle number density $\rho$ and (b) enthalpy per particle $\hat h$  as functions of $T$ for a given $p$. 
}
\label{fig:T-HV}
\end{center}
\end{figure}

Let us start with a familiar liquid--gas transition observed in a fluid at constant pressure $p$ in contact with a heat bath of temperature $T$.
The density of the fluid exhibits a discontinuous jumps at $T=\Tc(p)$.
The fluid behaves as a liquid at $T<\Tc(p)$ and as gas at $T>\Tc(p)$.
Letting the equation of state for the liquid and gas be $\rho=\rho^\subL(T,p)$ and $\rho=\rho^\subG(T,p)$, respectively, 
we obtain the particle number density at the liquid--gas transition as 
$\rho_\subC^\subL(p)=\lim_{T\rightarrow \Tc(p)^{-}}\rho^\subL(T,p)$
and
$\rho_\subC^\subG(p)=\lim_{T\rightarrow \Tc(p)^{+}}\rho^\subG(T,p)$.
See Fig.~\ref{fig:T-HV}(a).
The latent heat  generated in a transition to gas from liquid is
equivalent  to the jump in enthalpy $H$, where $H=U+pV$ with the  internal energy $U$ and the volume $V$ of the system.
As shown in Fig.~\ref{fig:T-HV}(b), letting $\hat h^\subL(T,p)$ and $\hat h^\subG(T,p)$ be the enthalpy per particle for liquid and gas, the latent heat per particle is
\begin{align}
\hat q(p)=\hat h_\subC^\subG(p)-\hat h_\subC^\subL(p),
\end{align}
where 
$\hat h_\subC^\subL(p)=\lim_{T\rightarrow \Tc(p)^{-}}\hat h^\subL(T,p)$ 
and
$\hat h_\subC^\subG(p)=\lim_{T\rightarrow \Tc(p)^{+}}\hat h^\subG(T,p)$. 

\begin{figure}[bt]
\begin{center}
\includegraphics[scale=0.43]{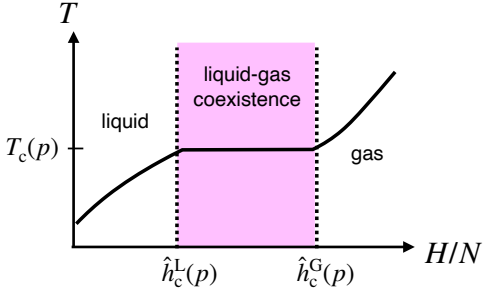}
\caption{Temperature as a function of $\hat h=H/N$ for equilibrium systems with $p$ fixed. The plateau corresponds to the transition temperature $T_\subC(p)$. Swapping the vertical and horizontal axes of the graph in Fig. ~\ref{fig:T-HV}(b) yields this figure.}
\label{fig:eos}
\end{center}
\end{figure}

For a thermally isolated fluid from the environment of constant pressure,
enthalpy $H$ is the conserved energy, which corresponds to the sum of the energy of the system and the interaction potential between the system and the environment.
Figure \ref{fig:eos} shows the equilibrium temperature of the system as a function of $\hat h=H/N$ with $p$ fixed. 
This figure is equivalent to Fig.~\ref{fig:T-HV}(b) except for the swapping of the vertical and horizontal axes.
The system exhibits stable liquid--gas coexistence for a certain range of enthalpy, say $ \hat h_\subC^\subL(p)<\hat h< \hat h_\subC^\subG(p)$, while it is occupied by liquid for $\hat h< \hat h_\subC^\subL(p)$ or by gas for $\hat h >\hat h_\subC^\subG(p)$.
Choosing $\hat h$ as a certain value in the range of coexistence, the distribution of $H$ and $N$ to the liquid and the gas are uniquely determined as $H^\subL$ with $H^\subG=H-H^\subL$ and $N^\subL$ with $N^\subG=N-N^\subL$.

A remarkable point in equilibrium thermodynamics is the correspondence of the thermodynamic entropy $S$ with the entropy as the variational function $\mathscr S$ in their functional forms.
Suppose that ${\cal H}^\subL$ and ${\cal N}^\subL$ are the variational variables corresponding to the distribution of  the enthalpy and the number of particles to liquid,
which should be uniquely determined in equilibrium as $H^\subL$ and $N^\subL$.
The entropy function as the variational function is given by
\begin{align}
{\mathscr S}({\cal H}^\subL,{\cal N}^\subL;H,p,N)=S({\cal H}^\subL,p,{\cal N}^\subL)+S({\cal H}^\subG,p,{\cal N}^\subG)
\label{e:Svar-eq}
\end{align}
with  ${\cal H}^\subG=H-{\cal H}^\subL$ and  ${\cal N}^\subG=N-{\cal N}^\subL$.
Meanwhile, the thermodynamic entropy $S$ of the total system satisfies
\begin{align}
S(H,p,N)=S(H^\subL,p,N^\subL)+S(H^\subG,p,N^\subG),
\label{e:Sth-eq}
\end{align}
where $H^\subG=H-H^\subL$ and $N^\subG=N-N^\subL$.
The functional form of the right-hand side of \eqref{e:Sth-eq} is the same as that of \eqref{e:Svar-eq}.
However, we note the difference in variables between $\mathscr S$ and $S$;
i.e., $\mathscr S$ is a five-variable function whereas $S$ is a three-variable function. The two functions appear in different contexts. 

\begin{figure}[bt]
\begin{center}
\includegraphics[scale=0.43]{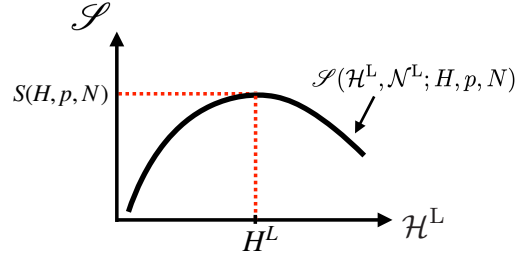}
\caption{Schematic figure for the entropy $\mathscr S$ as the variational function. The axis of ${\cal N}^\subL$ is omitted for clarity.}
\label{fig:var}
\end{center}
\end{figure}

In equilibrium thermodynamics, thermodynamic entropy $S$ is related to the variational function $\mathscr S$ as follows.
Equilibrium states maximize the entropy (variational function) $\mathscr S$ according to the second law of thermodynamics, and the equilibrium states are  expressed as
\begin{align}
  &(H^\subL, N^\subL)
  =\argmax_{{\cal H}^\subL, ~{\cal N}^\subL}
    {\mathscr S}({\cal H}^\subL,{\cal N}^\subL;H,p,N).
\label{e:var-eq}
\end{align}
Thus,  $H^\subL$ and $N^\subL$ are the solutions to the simultaneous equations for ${\cal H}^\subL$ and ${\cal N}^\subL$,
\begin{align}
&\left.\pderf{\mathscr S}{{\cal H}^\subL}{N^\subL,H,p,N}\right|_{{\cal H}^\subL=H^\subL}=0,\label{e:var-eq-HL}\\
&\left.\pderf{\mathscr S}{{\cal N}^\subL}{H^\subL,H,p,N}\right|_{{\cal N}^\subL=N^\subL}=0,\label{e:var-eq-NL}
\end{align}
from which we identify the equilibrium state as
\begin{align}
H^\subL=H^\subL(H,p,N), \quad N^\subL=N^\subL(H,p,N).
\end{align}
Substituting this solution into \eqref{e:Svar-eq}, 
we obtain
\begin{align}
{\mathscr S}(H^\subL(H,p,N), N^\subL(H,p,N);H,p,N)= S(H,p,N),
\end{align}
which means that the maximum value of the variational function $\mathscr S$
is equal to the thermodynamic entropy $S$. We also have 
\begin{align}
S(H,p,N)=&S(H^\subL(H,p,N), p,N^\subL(H,p,N))\nonumber\\
&+S(H^\subG(H,p,N),p,N^\subG(H,p,N))
\end{align}
as the exact expression of \eqref{e:Sth-eq}.
The relation of $S$ with $\mathscr S$ is presented in Fig.~\ref{fig:var}.

The temperatures of the liquid and gas are defined as
\begin{align}
\frac{1}{T^\subL}\equiv \pderf{S}{H^\subL}{p,N^\subL},\quad
\frac{1}{T^\subG}\equiv \pderf{S}{H^\subG}{p,N^\subG}.
\end{align}
Substituting \eqref{e:Svar-eq} with ${\cal H}^\subG=H-{\cal H}^\subL$ into the variational equation \eqref{e:var-eq-HL},
we have
\begin{align}
T^\subL=T^\subG,
\end{align}
which is the balance of temperature between the liquid and gas in equilibrium.
Similarly, because the chemical potentials are given by
\begin{align}
\frac{\mu^\subL}{T^\subL}\equiv -\pderf{S}{N^\subL}{p,H^\subL},\quad
\frac{\mu^\subG}{T^\subG}\equiv -\pderf{S}{N^\subG}{p,H^\subG},
\end{align}
the variational equation \eqref{e:var-eq-NL} together with $T^\subL=T^\subG$ and ${\cal N}^\subG=N-{\cal N}^\subL$ yields 
\begin{align}
\mu^\subL=\mu^\subG, 
\end{align}
which corresponds to the balance of the chemical potential between the liquid and gas.
With these two balances,  we obtain the fundamental relation of thermodynamics for the total system as
\begin{align}
dS=\frac{1}{T}dH-\frac{V}{T}dp-\frac{\mu}{T}dN,
\end{align}
in which we write $T=T^\subL=T^\subG$ and $\mu=\mu^\subL=\mu^\subG$. 
$V=V^\subL+V^\subG$ is the total volume of the system.

As we have reviewed so far,  the correspondence of the two entropy functions, $\mathscr S$ and $S$, is concluded from the second law of thermodynamics,
and the correspondence  is  a strong property of the equilibrium entropy
describing macroscopic phenomena.


\section{Setup}
\label{s:setup}

We focus on enthalpy-conserving systems at constant pressure in heat conduction.  See Fig.~\ref{fig:setup}.
$N$ particles are enclosed in a container made of thermally insulating
material except for the left and right boundaries.
The top plate of the container at the position $L_z$ is set freely movable to keep the pressure constant, 
whereas the other five walls are fixed with constant lengths $L_x$ and $L_y$.
The volume of the system $V=L_xL_yL_z$ is not fixed whereas the total enthalpy $H$ of the system is kept constant.
We assume that the transition temperature $\Tc(p)$ is far below the critical temperature.
For simplicity, gravity effects are ignored such that the pressure is uniform over the system.

\begin{figure}[t]
\begin{center}
\includegraphics[scale=0.34]{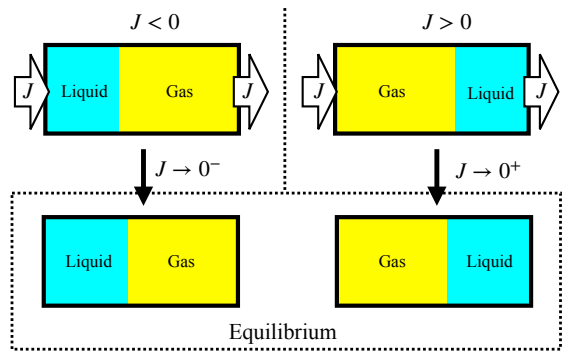}
\caption{Violation of left--right symmetry in heat conduction. The steady state at $J<0$ is symmetric to that at $J>0$. Both mirror-image states become stable exactly at $J=0$.}
\label{fig:symmetry}
\end{center}
\end{figure}

To conserve enthalpy, we control the in- and out-flow heat currents from the left and right boundaries to be strictly $J L_y L_z$ per unit time.
For $J<0$, the liquid localizes on the left side of the system,
and there is an interface that separates the liquid from the gas. 
Such violation of left--right symmetry continues to $J\rightarrow 0^-$.
The steady state for $J >0$ is obtained by the space-inversion of the system with $-J$ as shown in Fig.~\ref{fig:symmetry}. When $J$ is continuously changed from negative to positive values, local thermodynamic quantities may jump from those of liquid to gas or vise versa at $J=0$. This indicates the singularity at $J=0$.

For fixed values of $H$, $p$, $N$, and $J$, the enthalpy and particle  distributions are determined uniquely.
We assume that $|J|$ is so small that  each local thermodynamic quantity is uniform in the section perpendicular to $x$, and we therefore write the local quantity as a function of $x$.
This means that the heat conduction system remains extensive in $y$ and $z$ directions even for $J\neq 0$ because the value of $L_y$ or $L_z$ does not affect any local state.
Using local temperature $T(x)$,
we write the temperatures at the two ends as 
\begin{align}
T_1=\lim_{x\rightarrow 0^+}T(x), \quad T_2=\lim_{x\rightarrow L_x^{-}}T(x).
\end{align}
We observe liquid--gas coexistence when  $T_1<\Tc(p)<T_2$ for $J<0$, or $T_1>\Tc(p)>T_2$ for $J>0$.
As a dimensionless parameter representing the degree of nonequilibrium, we set
\begin{align}
\ep=\frac{T_2-T_1}{\Tc(p)}.
\label{e:ep}
\end{align}

Suppose that the liquid--gas interface is situated at $x=X$,
where  the thickness of the interface is not macroscopic and is ignored in the present description.
We express the enthalpy and the particle number
 of the phase in the left (right) region of $x<X$ ($x>X$) as
\begin{align}
&H^\subl=L_y L_z\intL h(x),\quad H^\subr=L_y L_z\intG h(x),\\
&N^\subl=L_y L_z\intL \rho(x),\quad N^\subr=L_y L_z\intG \rho(x),
\end{align}
where $h(x)$ and $\rho(x)$ denote  local enthalpy density and number density.
The total enthalpy and total number of particles in the system are
\begin{align}
H=H^\subl+H^\subr, \quad N=N^\subl+N^\subr.
\end{align}
Conventionally, one may
  interpret $\subl=\subL$ and $\subr=\subG$ for $J <0$
  and vise versa for  $J >0$.  The variational variables ${\cal H}^\subLG$ and  
  ${\cal N}^\subLG$ in Sec. \ref{s:review} are replaced with 
  ${\cal H}^\sublr$ and  ${\cal N}^\sublr$.

When we know the values of $T(x)$ and $p$ and the local phase as liquid or gas, we can identify any local thermodynamic quantity from equilibrium thermodynamics regardless of whether the local state is stable or metastable.
For instance, when the local phase is liquid, local entropy density is written as $s(x)=\rho^\subL(T(x), p) \hat s^\subL(T(x),p)$ with equilibrium entropy $\hat s(T,p)$ per particle.
The entropies for the left (right) region are expressed as the spatial integration of the local density; i.e.,
\begin{align}
S^\subl=L_y L_z\intL s(x),\quad
S^\subr=L_y L_z\intG s(x).
\label{e:S-LG}
\end{align}
These entropies are defined in heat conduction; however they are written by the equilibrium entropy functions,
\begin{align}
S^\subl=S(H^\subl,p,N^\subl), \quad
S^\subr=S(H^\subr,p,N^\subr)
\label{e:S-LG-th}
\end{align} 
with an error of $O(\ep^2)$, as shown  in Sec.3 of \cite{global-JSP}.
See Appendix \ref{s:GTD-LG}.

In the last of this section, we mention the treatment of the singularity at $J=0$ associated with the left--right symmetry demonstrated in Fig.~\ref{fig:symmetry}.
When $J$ is varied continuously from negative to positive values, thermodynamic quantities for the left region become discontinuous at $J=0$. We assume that this discontinuity can be expressed as the limit of smooth functions of $\ep$ by introducing some regularization parameters.
 We set the regularization parameters
 to be zero in the last step of the calculation of the thermodynamic quantities. Here, we choose the regularization parameters in such a way that the global thermodynamics is formulated consistently. 

\section{Problem}
\label{s:problem}

Our aim in this paper is to extend the maximum entropy principle
to heat conduction states.
We introduce a nonequilibrium intensive variable $\phi$ and its conjugate
extensive variable $\Psi$ with the condition
\begin{align}
\phi=O(\ep), \quad \Psi=O(\ep^0).
\end{align}
We then seek extended forms of the entropy $S$ and
variational function ${\mathscr S}$ that satisfy 
three conditions.

The first condition is that $S$ and ${\mathscr S}$
take the same form 
\begin{align}
&S(H,p,N,\phi)=S(H^\subl,p,N^\subl)+S(H^\subr,p,N^\subr)\nonumber\\
&~~~~~~~~~~~~~~~~~~~~~~~~~~~~~+\phi\Psi(H,p,N,\phi), \label{e:Sth-neq}\\
&{\mathscr S}({\cal H}^\subl,{\cal N}^\subl;H,p,N,\phi)=
S({\cal H}^\subl,p,{\cal N}^\subl)\nonumber\\
&~~~~~~~~~+S({\cal H}^\subr,p,{\cal N}^\subr)+\phi {\psi}({\cal H}^\subl,{\cal N}^\subl;H,p,N), 
\label{e:Svar-neq}
\end{align}
where the function $\psi$ does not depend on $\phi$ explicitly.

The second condition for $S$ and ${\mathscr S}$ is that the
thermodynamic entropy $S$
corresponds to the maximum value of the variational function $\mathscr S$; i.e.,
\begin{align}
&S(H,p,N,\phi)\nonumber\\
&~={\mathscr S}(H^\subl(H,p,N,\phi),N^\subl(H,p,N,\phi);H,p,N,\phi),
\label{e:S*def}
\end{align}
where the function $\psi$ is connected to $\Psi$ as
\begin{align}
\Psi(H,p,N,\phi)=\psi(H^\subl,N^\subl;H,p,N)
\label{e:Psi*def}
\end{align}    
with the steady state values
\begin{align}
H^\subl=H^\subl(H,p,N,\phi), \quad N^\subl=N^\subl(H,p,N,\phi).
\label{e:steadyHN}
\end{align}
We assume that the steady state maximizes $\mathscr S$. Then, $H^\subl$ and $N^\subl$  are determined as
\begin{align}
  &(H^\subl, N^\subl)
  =\argmax_{{\cal H}^\subl, ~{\cal N}^\subl}
    {\mathscr S}({\cal H}^\subl,{\cal N}^\subl;H,p,N,\phi).
\label{e:var-neq}
\end{align}

The third  condition for $S$ and ${\mathscr S}$ is that $S(H,p,N,\phi)$
satisfies the fundamental relation of thermodynamics 
\begin{align}
dS=\frac{1}{\bT}dH-\frac{V}{\bT}dp-\frac{\bmu}{\bT}dN+\Psi d\phi,
\label{e:1st-law}
\end{align}
where $\bT$ is the global temperature defined by
\begin{align}
\bT=\frac{\intx \rho(x)T(x)}{\intx \rho(x)},
\label{e:bT}
\end{align}
which was introduced in \cite{global-PRL}. 
We also define the global chemical potential as
\begin{align}
\bmu=\frac{\intx \rho(x) \mu(x)}{\intx \rho(x)},
\label{e:bmu}
\end{align}
where $\mu(x)$ is the local chemical potential.
We emphasize that, once we decide to adopt $\bT$ as a temperature to characterize the heat conduction states, $\bmu$ is concluded to be a unique extension of the equilibrium chemical potential. See Sec. 7 of \cite{global-JSP}.

Summarizing the above, we attempt to determine an entropy extended to liquid--gas coexistence in heat conduction  using simultaneous equations, specifically
two equations from the variational principle \eqref{e:var-neq},
\begin{align}
&\left.\pderf{\mathscr S}{{\cal H}^\subl}{N^\subl,H,p,N,\phi}\right|_{{\cal H}^\subl=H^\subl}=0,\label{e:var-neq-HL}\\
&\left.\pderf{\mathscr S}{{\cal N}^\subl}{H^\subl,H,p,N,\phi}\right|_{{\cal N}^\subl=N^\subl}=0,\label{e:var-neq-NL}
\end{align}
and four equations from the fundamental relation \eqref{e:1st-law},
\begin{align}
&\pderf{S}{H}{p,N,\phi}=\frac{1}{\bT},\label{e:th-neq-H}\\
&\pderf{S}{p}{H,N,\phi}=-\frac{V}{\bT},\label{e:th-neq-p}\\
&\pderf{S}{N}{H,p,\phi}=-\frac{\bmu}{\bT},\label{e:th-neq-N}\\
&\pderf{S}{\phi}{H,p,N}=\Psi.\label{e:th-neq-phi}
\end{align}

\section{Results}
\label{s:results}

Solving the equations \eqref{e:var-neq-HL}, \eqref{e:var-neq-NL}, \eqref{e:th-neq-H}, \eqref{e:th-neq-p}, \eqref{e:th-neq-N}, and \eqref{e:th-neq-phi} 
under the conditions described in Sec. \ref{s:setup},
we determine  the functional form of $\phi$ and $\psi$  as 
\begin{align}
\phi=\ep
\label{e:phi}
\end{align}
and
\begin{align}
\psi({\cal H}^\subl,{\cal N}^\subl;H,p,N)=
\frac{{\cal H}^\subr {\cal N}^\subl-{\cal H}^\subl {\cal N}^\subr}{2NT_\subC(p)}
\label{e:Psi}
\end{align}
up to a multiplicative constant. This multiplicative constant does not affect
any quantitative prediction and we thus set it as a specific value as discussed in Appendix \ref{sec:det-psi}

The steady-state value of $\psi$ is expressed as
\begin{align}
  \Psi(H,p,N,\phi)=
 \frac{\phi}{|\phi|}
  \frac{\hat q(p)}{2T_\subC(p)}\frac{N^\subL N^\subG}{N}
\label{e:Psi2}
\end{align}
using \eqref{e:Psi*def} and \eqref{e:Psi},
where $N^\subLG$ is the steady number of particles in the liquid and gas
and we have used the relation for the steady state value of enthalpy as
$H^\subLG=\hat h_\subC^\subLG(p) N^\subLG+O(\ep)$.
The expression \eqref{e:Psi2} clarifies that the new quantity $\Psi$
is associated with the phase coexistence
because it is connected to the latent heat $\hat q(p)$.
Meanwhile, $\phi$ is a nonequilibrium variable that is never associated with equilibrium properties.
Moreover, we emphasize that $\phi$ is uniquely determined by  \eqref{e:phi} as the difference in temperature at the two boundaries and not as the steady heat current $J$ although $J$ is an operational parameter that controls the enthalpy conservation in heat conduction.

Using the variational function \eqref{e:Svar-neq} with \eqref{e:phi}
and \eqref{e:Psi}, the variational equations \eqref{e:var-neq-HL}
and \eqref{e:var-neq-NL} are expressed as
\begin{align}
&\frac{1}{\bT^\subl}-\frac{1}{\bT^\subr}=\frac{\phi}{2T_\subC(p)},\label{e:var-sol-HL}\\
&\frac{\bmu^\subl}{\bT^\subl}-\frac{\bmu^\subr}{\bT^\subr}=\frac{\phi}{2T_\subC(p)}\frac{H}{N}.\label{e:var-sol-NL}
\end{align}
Here, $(\subl,\subr)=(L,G)$ for $\ep >0$ and  $(\subl,\subr)=(G,L)$ for $\ep <0$,
where $\bT^\subL$ and $\bT^\subG$ are global temperatures for the liquid and gas, respectively, 
and $\bmu^\subL$ and $\bmu^\subG$ are the respective global chemical potentials.
\eqref{e:var-sol-HL} and \eqref{e:var-sol-NL} at $\ep=0$ correspond to the equilibrium balances of temperature and chemical potential between the liquid and gas.
Thus, these relations at $\ep\neq 0$ provide the first-order correction of the balances.
Indeed, \eqref{e:var-sol-HL} is transformed as
\begin{align}
    \bT^\subr-\bT^\subl=\frac{\phi}{2} T_\subC(p)+O(\ep^2),
\end{align}    
which leads to \eqref{e:ep} with $\phi=\ep$ as derived from \eqref{e:bT-LG}.

We emphasize that \eqref{e:var-sol-HL} and \eqref{e:var-sol-NL}  are sufficient to identify steady states.
As we will demonstrate in Sec. \ref{s:steady-state},
\eqref{e:var-sol-HL} and \eqref{e:var-sol-NL} lead to all thermodynamic quantities for the  liquid--gas coexistence in heat conduction.
For instance, \eqref{e:theta-J} and \eqref{e:Xneq-x} provide experimentally accessible relations for the steady states, which may act as the check of the validity of the present theory. 
We will also see that $\theta$ is not necessarily equal to $\Tc(p)$ because metastable states, such as super-cooled gas or super-heated liquid, may be stably observed in heat conduction.
Moreover, the steady states are consistent with those determined by the variational principle for heat conduction systems attached to two heat baths of $T_1$ and $T_2$ \cite{global-JSP}.
See Sec. \ref{s:equivalence} for the correspondence between the enthalpy-conserving and nonconserving systems.

\section{Thermodynamic equivalence of steady states in heat conduction}\label{s:equivalence}
\label{s:equiv}

\begin{figure}[tb]
\begin{center}
\includegraphics[scale=0.45]{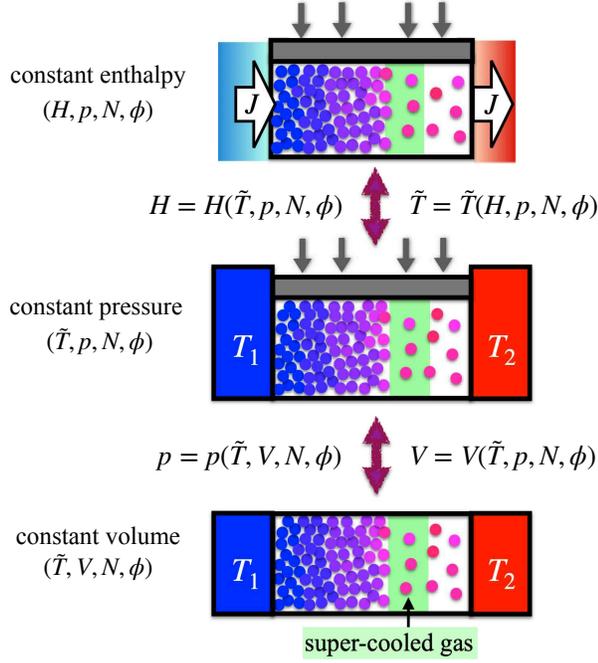}
\caption{Thermodynamic equivalence of heat conduction systems.}
\label{fig:equivalence}
\end{center}
\end{figure}

The thermodynamic relations of equilibrium systems are independent of
whether the system is adiabatic or isothermal and whether the system is at
constant pressure or constant volume. Remarkably, the equivalence
between the fundamental thermodynamic relations can be formulated
using the Legendre transformation. We then ask whether there 
exist similar relations among different formulations for heat conduction
systems.  See Table \ref{t:summary}.  Specifically, we study the three
cases presented in Fig. \ref{fig:equivalence}. So far, we have studied 
enthalpy conserving systems at constant pressure in heat conduction,
which is shown in the top panel. In the second panel, we attach a heat
bath of $T_1$ at the left wall and
a heat bath of $T_2$ at the right wall, where $T_1 \le T_2$ is assumed without
loss of generality. 
Below, we assume that there are no temperature gaps between the system
and the heat baths (i.e., $T(0^+)=T_1$ and $T(L_x^-)=T_2$) in steady
states. In the bottom panel of Fig. \ref{fig:equivalence}, the system
is kept at constant volume instead of at constant pressure. 
While the fundamental thermodynamic relations and the variational principles for the systems of constant pressure or constant volume were investigated in \cite{global-JSP},
we reformulate them from the results in the previous section.

\begin{table*}[tb]
\begin{tabular}{|c|c|c|}\hline
 & Variational function  &Thermodynamic relation \\\hline
\begin{tabular}{c} constant enthalpy \\ $(H,p,N,\phi)$ \end{tabular} & ~${\mathscr S}=L_y L_z \int_0^{L_x} s(x)~dx +\phi\psi$~  &~${\displaystyle dS=\frac{dH}{\bT}-\frac{V}{\bT}dp-\frac{\bmu}{\bT}dN+\Psi d\phi}$~ \\\hline
\begin{tabular}{c} constant pressure \\ $(\bT,p,N,\phi)$ \end{tabular}  & ~${\cal G}=L_y L_z \int_0^{L_x}  \mu(x)\rho(x)~dx $~  & ~${\displaystyle dG=-Sd\bT+{V}dp+{\bmu}dN-\bT\Psi d\phi}$~ \\\hline
\begin{tabular}{c} constant volume \\ $(\bT,V,N,\phi)$ \end{tabular}  & ~${\cal F}=L_y L_z \int_0^{L_x} f(x)~dx $~ &~${\displaystyle dF=-Sd\bT-pdV+{\bmu}dN-\bT\Psi d\phi}$~ \\\hline
\end{tabular}
\caption{Summary of variational functions and fundamental relations of thermodynamics. $S$ is the maximum value of $\mathscr S$ whereas $G$ and $F$ are the minimum values of $\cal G$ and $\cal F$, respectively. The fundamental relations are associated with one another via a Legendre transformation.}
\label{t:summary}
\end{table*}

\subsection{Thermodynamic relations}

We define the Gibbs free energy in heat conduction as
\begin{align}
G=L_yL_z\intx \mu(x)\rho(x).
\label{e:bG-def}
\end{align}
The entropy $S$ given in \eqref{e:Sth-neq} with
$\phi=\ep$ and $\Psi$ as \eqref{e:Psi2} satisfies
\begin{align}
S=\frac{H-G}{\bT}+O(\ep^2)
\label{e:LeGendre}
\end{align}
with the global temperature $\bT$ in \eqref{e:bT}. 
See Appendix \ref{s:LeGendre} for the derivation of \eqref{e:LeGendre}.
Substituting \eqref{e:LeGendre} into \eqref{e:1st-law}, we obtain
\begin{align}
dG=-Sd\bT+Vdp+\bmu dN-\bT\Psi d\phi.
\label{e:1st-lawG}
\end{align}
From this relation, we find that the functional form
\begin{align}
G=G(\bT,p,N,\phi)
\end{align}
is useful. 
Because the global Helmholtz free energy
\begin{align}
F=L_y L_z\intx f(x)
\label{e:bF-def}
\end{align}
satisfies
\begin{align}
F=G-pV,
\end{align}
\eqref{e:1st-lawG} is transformed into the fundamental relation 
\begin{align}
dF=-Sd\bT-pdV+\bmu dN-\bT\Psi d\phi,
\label{e:1st-lawF}
\end{align}
which leads to the expression
\begin{align}
F=F(\bT,V,N,\phi).
\end{align}
Thus, the fundamental relations of global thermodynamics are formed consistently among $(H,p,N,\phi)$, $(\bT,p,N,\phi)$, and $(\bT,V,N,\phi)$. 
The transformation into $(U,V,N,\phi)$ is also straightforward.
These transformations are nothing but the Legendre transitions.
See Appendix \ref{s:add-GF} for further arguments on the additivity of $G$ and $F$.

We here present a remark.
The fundamental relations of global thermodynamics derived in \cite{global-JSP}
were expressed in slightly different forms from \eqref{e:1st-lawG} and \eqref{e:1st-lawF}; i.e.,
\begin{align}
dG=-(S^\subL+S^\subG)d\bT+Vdp+\bmu dN-\Psi d\Xi,
\label{e:Gibbs2}
\\
dF=-(S^\subL+S^\subG)d\bT-pdV+\bmu dN-\Psi d\Xi,
\label{e:GibbsF2}
\end{align}
where $\Xi=T_2-T_1$.
\eqref{e:Gibbs2} and \eqref{e:GibbsF2}  are obtained by substituting $\phi=\Xi/\bT$
into \eqref{e:1st-lawG} and \eqref{e:1st-lawF}.
While the expression \eqref{e:Gibbs2} is equivalent to \eqref{e:1st-lawG}, the former implies the extension of the Gibbs free energy in the form $G=G(\bT,p,N,\Xi)$,
in which the nonequilibrium variable is not $\phi$ but $\Xi$.
As long as we discuss thermodynamic properties for the systems attached to
heat baths, both the relations work consistently. However, the connection
to the thermodynamic relation in enthalpy-conserving systems
is more clearly seen in \eqref{e:1st-lawG}.

\subsection{Variational principles}

We first explain the variational principle corresponding to the minimization of the Gibbs free energy $\cal G$. In equilibrium, this minimization is performed by fixing $(T,p,N)$. 
The extension of this variational principle to heat conduction is not straightforward because the temperature profile cannot be given as an external parameter of the setup. 
Nevertheless, referring to the thermodynamic relation \eqref{e:1st-lawG}, 
we formally adopt the fixed condition as $(\bT, p, N,\phi)$.
To fix $\bT$ and $\phi$ in the variation at constant $p$,  
we shift the two temperatures by the same amount while fixing $T_2-T_1(=\Xi)$
and find heuristically the shift required to keep the right-hand side of \eqref{e:bT}
as $\bT$.
Thus, fixing $(\bT,p,N,\phi)$ is the same as fixing $(\bT,p,N,\Xi)$, whose variational principle and the steady states given by it have already been investigated in \cite{global-JSP}.

In brief, taking a variational function as a natural extension from equilibrium, 
\begin{align}
{\cal G}=L_yL_z\intx \mu(x)\rho(x),
\end{align}
we obtain
\begin{align}
&{\cal G}({\cal N}^\subL;\bT,p,N,\phi)=G^\subL({\cal T}^\subL({\cal N}^\subL;\bT,N,\phi),p,{\cal N}^\subL)\nonumber\\
&~~~~~
+G^\subG({\cal T}^\subG({\cal N}^\subL;\bT,N,\phi),p,N-{\cal N}^\subL),
\label{e:Gvar-neq}
\end{align}
where
$G^\subL({\cal T}^\subL, p, {\cal N}^\subL)=\mu^\subL({\cal T}^\subL,p){\cal N}^\subL$
and $G^\subG({\cal T}^\subG,p,{\cal N}^\subG)=\mu^\subG({\cal T}^\subG,p)(N-{\cal N}^\subG)$.
 ${\cal T}^\subL$ and ${\cal T}^\subG$ are the global temperatures for the liquid and gas, which are considered as variational quantities because they depend on ${\cal N}^\subL$ according to
\begin{align}
{\cal T}^\subL=\bT\left(1-\phi \frac{N-{\cal N}^\subL}{2N}\right),\quad
{\cal T}^\subG=\bT\left(1+\phi\frac{{\cal N}^\subL}{2N}\right).
\label{e:var-TLTG}
\end{align}
\eqref{e:var-TLTG} results from the definition of $\bT$ and $\phi$ with $\bT=T_\subC(p)+O(\ep)$
assuming that local temperature is continuous everywhere, even at the liquid--gas interface. 
See the discussion around \eqref{e:TG-TL}.

The steady state is specified by determining the distribution of particles to liquid as
\begin{align}
N^\subL=\argmin_{{\cal N}^\subL} {\cal G}({\cal N}^\subL;\bT,p,N,\phi),
\label{e:var-G}
\end{align}
which is rewritten as 
\begin{align}
\left.\pderf{\cal G}{{\cal N}^\subL}{\bT,p,N,\phi}\right|_{{\cal N}^\subL=N^\subL}=0.
\label{e:var-G-NL}
\end{align}
Using \eqref{e:Gvar-neq} together with \eqref{e:var-TLTG}, the variational equation \eqref{e:var-G-NL} leads to
\begin{align}
\mu^\subL-\mu^\subG=\phi \bT\frac{S^\subL+S^\subG}{2N},
\label{e:muL-muG}
\end{align}
where we applied the fundamental relation of global thermodynamics, which is  briefly reviewed in Appendix \ref{s:GTD-LG}.
We show the equivalence between \eqref{e:var-sol-NL} and \eqref{e:muL-muG} in Appendix \ref{s:temp-relation}.

We next deal with constant-volume systems in heat conduction.
We assume that the steady state for a given $(\bT, V, N,\phi)$ is obtained as the state minimizing the total Helmholtz free energy of the system, 
\begin{align}
{\cal F}=L_y L_z \intx f(x).
\end{align}
More precisely, we formulate the variational function as
\begin{align}
&{\cal F}({\cal V}^\subL, {\cal N}^\subL;\bT,V,N,\phi)=
F({\cal T}^\subL({\cal N}^\subL;\bT,N,\phi),{\cal V}^\subL,{\cal N}^\subL)\nonumber\\
&~~~~~
+F({\cal T}^\subG({\cal N}^\subL;\bT,N,\phi),V-{\cal V}^\subL,N-{\cal N}^\subL),
\end{align}
where $\cal T^\subL$ and $\cal T^\subG$ are functions given by \eqref{e:var-TLTG}.
The steady state is obtained as 
\begin{align}
(V^\subL,N^\subL)=\argmin_{{\cal V}^\subL,~{\cal N}^\subL} {\cal F}({\cal V}^\subL,{\cal N}^\subL;\bT,V,N,\phi).
\label{e:var-F}
\end{align}
That is to say,
\begin{align}
&\left.\pderf{\cal F}{{\cal V}^\subL}{\bT,V,N,\phi,N^\subL}\right|_{{\cal V}^\subL=V^\subL}=0,\label{e:var-F-VL}\\
&\left.\pderf{\cal F}{{\cal N}^\subL}{\bT,V,N,\phi,V^\subL}\right|_{{\cal N}^\subL=N^\subL}=0.
\label{e:var-F-NL}
\end{align}
The first equation \eqref{e:var-F-VL} leads to
\begin{align}
p^\subL=p^\subG
\end{align}
corresponding to the balance of pressure in the liquid and gas.
The second equation \eqref{e:var-F-NL} yields \eqref{e:muL-muG}.

In summary,  the minimization of free energy $\cal G$ with $(\bT,p,N,
\phi)$ fixed and the minimization of  $\cal F$ with $(\bT,V,N,\phi)$ fixed 
provide the same results as  the maximization of entropy $\mathscr S$
with $(H,p,N,\phi)$ fixed; i.e., the steady states are equivalent among the three.

\section{Steady states}
\label{s:steady-state}

\begin{figure}[bt]
\begin{center}
\includegraphics[scale=0.35]{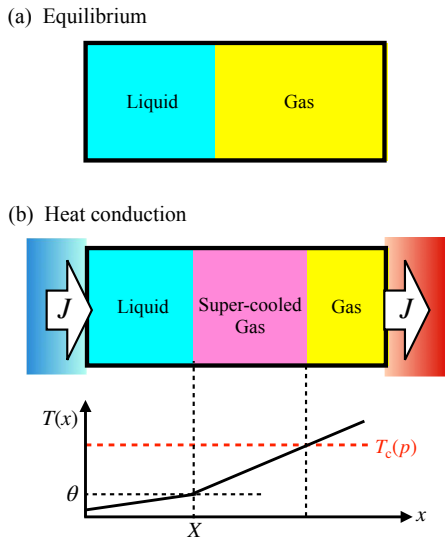}
\caption{
Schematic figures comparing a heat conduction state with an equilibrium state.
(a) Equilibrium state at $J\rightarrow 0^-$. Local temperature is $T_\subC(p)$ at any $x$.
(b) Heat conduction state  with $J<0$ and the temperature profile $T(x)$. 
Super-cooled gas occupies a rather wide region.}
\label{fig:ss}
\end{center}
\end{figure}

This section summarizes the thermodynamic quantities in the steady heat conduction state, which result from the variational principle. The derivations of the formulas are presented in Appedix \ref{s:steady-state-derive}.

In Sec. \ref{s:theta}, we refer to the expression of the temperature at the liquid--gas interface, which was derived in \cite{global-JSP}.
We demonstrate 
various quantities in the steady state of the enthalpy-conserving systems with $J\neq 0$
as slight changes from corresponding equilibrium quantities in Sec. \ref{s:SS-HPN}.
Such perturbative treatment is difficult to apply to systems in contact with two heat baths at constant pressure
because the equilibrium position of the liquid--gas interface is not unique at $T_\subC(p)$, 
whereas the liquid and gas do not coexist at $T\neq T_\subC(p)$.
This indicates that we cannot formulate a naive perturbative expansion in heat conduction near equilibrium.
The thermodynamic equivalence holds in steady states as discussed
in Sec. \ref{s:equivalence}, and
we may therefore transform the formulas for the steady state in $(H,p,N,\phi)$ into $(\bT,p,N,\phi)$.
See Sec. \ref{s:ss-TPN}.
Below, we omit the dependence of $p$ for the equilibrium quantities at the transition temperature.

When we set $J$ as a parameter, the nonequilibrium variable $\phi=(T_2-T_1)/T_\subC$ is determined as
\begin{align}
\phi=-J\frac{L_x}{T_\subC}\frac{1}{\kappa_\subC^\subL\kappa_\subC^\subG}
\frac{
\kappa_\subC^\subL\rho_\subC^\subL(\hat h- \hat h_\subC^\subL)
+
\kappa_\subC^\subG\rho_\subC^\subG(\hat h_\subC^\subG-\hat h)
}
{\rho_\subC^\subL(\hat h-\hat h_\subC^\subL)+\rho_\subC^\subG(\hat h_\subC^\subG-\hat h)}+O(\ep^2),
\label{e:phi-J}
\end{align}
where
$\kappa_\subC^\subLG$ is heat conductivity. 
Note that \eqref{e:phi-J} holds regardless of the sign of $J$ and  $\phi J\le 0$. 
In the estimates below, we use $\phi$ as it means \eqref{e:phi-J}.

A simple sketch of the steady state is presented in Fig.~\ref{fig:ss} for $J<0$ in comparison with equilibrium liquid--gas coexistence at $J\rightarrow 0^-$. There is a region occupied by super-cooled gas on the adjacent to the liquid--gas interface.
When we take $J>0$, the steady state is obtained as the mirror image of $J<0$ as shown in Fig.~\ref{fig:symmetry},
where the physical properties are equivalent to those for $J<0$.
Thus, all steady-state properties should be expressed by functions of $|J|$ or $|\phi|$.

\subsection{Temperature at the liquid--gas interface} \label{s:theta}

From the variational principle \eqref{e:var-sol-HL} and \eqref{e:var-sol-NL},
the temperature $\theta$ at the interface is formulated as
\begin{align}
&\theta={T_\subC}-
\frac{X(L_x-X)}{2L_x}
\left[
|J|\left(\frac{1}{\kappa_\subC^\subG}-\frac{1}{\kappa_\subC^\subL}
\right)
+\frac{|\phi|T_\subC}{L_x}\frac{\rho_\subC^\subL-\rho_\subC^\subG}{\bar\rho}
\right],
\label{e:theta-J}
\end{align}
in which $\bar \rho=N/V$ and $X$ is the width of the liquid region as reported in  
\cite{global-JSP}.
Another form of \eqref{e:theta-J} is 
\begin{align}
\theta=T_\subC-\frac{X(L_x-X)}{2L_x}
|J|\left(\frac{\rho_\subC^\subG}{\bar\rho}\frac{1}{\kappa_\subC^\subL}-\frac{\rho_\subC^\subL}{\bar\rho}\frac{1}{\kappa_\subC^\subG}\right)
\label{e:theta-J2}
\end{align}
as shown in Appendix \ref{s:theta-J}.
\eqref{e:theta-J} indicates that $\theta$ becomes lower than $T_\subC$ in the linear response regime.
Moreover,  the absolute values $|J|$ and $|\phi|$ in \eqref{e:theta-J} imply that
the true equilibrium state with a vanishing current $J=0$ is singular.
It is noted that $\theta$ may deviate from $T_c$ in experiments, as has been quantitatively discussed for a specific material such as water \cite{global-JSP}.

\subsection{Steady-state expressions as functions of $(H,p,N,\phi)$}
\label{s:SS-HPN}

From the variational equations \eqref{e:var-sol-HL} and \eqref{e:var-sol-NL},
we identify the width of the liquid $X=X(H,p,N,\phi)$, the volume $V=V(H,p,N,\phi)$, the number of  particles $N^\subL=N^\subL(H,p,N,\phi)$, and the enthalpy $H^\subL=H^\subL(H,p,N,\phi)$.
The following estimates have precision of $O(\ep)$ and error of $O(\ep^2)$.
The detailed derivation for each estimate is presented in Appendix \ref{s:steady-state-derive}.
 
In the equilibrium limit $|J|\rightarrow 0$, the interface corresponds to the Gibbs dividing surface. 
For a given $(H,p,N)$, we obtain the numbers of particles as
\begin{align}
&\frac{N_\eq^\subL}{N}=\frac{\hat h_\subC^\subG-\hat h}{\hat q},\quad
\frac{N_\eq^\subG}{N}=\frac{\hat h-\hat h_\subC^\subL}{\hat q},
\end{align}
the respective volumes as
\begin{align}
V_\eq^\subL=\frac{N}{\rho_\subC^\subL}\frac{\hat h_\subC^\subG-\hat h}{\hat q},\quad
V_\eq^\subG=\frac{N}{\rho_\subC^\subG}\frac{\hat h-\hat h_\subC^\subL}{\hat q},
\end{align}
and the width of the liquid as
\begin{align}
\frac{X_\eq}{L_x}=\frac{\rho_\subC^\subG(\hat h_\subC^\subG-\hat h)}{\rho_\subC^\subL(\hat h-\hat h_\subC^\subL)+\rho^\subG(\hat h_\subC^\subG-\hat h)}.
\label{e:xeq}
\end{align}

Imposing the heat current $J$, the distributions of conserved quantities to each bulk deviate as follows.
Letting $\hat c_p^\subLG$ be the specific heat at constant pressure,
the numbers of particles are
\begin{align}
&\frac{N^\subL-N_\eq^\subL}{N}=
-
|\phi|\frac{T_\subC}{2\hat q}
\left[
\left(\frac{\hat h_\subC^\subG-\hat h}{\hat q}\right)^2\hat c_p^\subL
-\left(\frac{\hat h -\hat h_\subC^\subL}{\hat q}\right)^2\hat c_p^\subG
\right],
\label{e:NL-pre}
 \\
&\frac{N^\subG-N_\eq^\subG}{N}=
|\phi|
\frac{T_\subC}{2\hat q}
\left[
\left(\frac{\hat h_\subC^\subG-\hat h}{\hat q}\right)^2\hat c_p^\subL
-\left(\frac{\hat h -\hat h_\subC^\subL}{\hat q}\right)^2\hat c_p^\subG
\right],
\label{e:NG-pre}
\end{align}
the enthalpies per particle are
\begin{align}
&\hat h^\subL-\hat h_\subC^\subL=-|\phi|\frac{T_\subC}{2}\frac{\hat h_\subC^\subG-\hat h}{\hat q}\hat c_p^\subL,
\label{e:HL-pre}
\\
&\hat h^\subG-\hat h_\subC^\subG=|\phi|\frac{T_\subC}{2}\frac{\hat h-\hat h_\subC^\subL}{\hat q}\hat c_p^\subG,
\label{e:HG-pre}
\end{align}
and the volumes are
\begin{align}
&V^\subL-V_\eq^\subL=
-|\phi|\frac{T_\subC}{2}
\frac{N}{\rho_\subC^\subL}\left[
\left(\frac{\hat h_\subC^\subG-\hat h}{\hat {q}}\right)^2
\alpha_\subC^\subL
\right.\nonumber\\
&~~~~~\left.
+
\left(
\left(\frac{\hat h_\subC^\subG-\hat h}{\hat q}\right)^2\frac{\hat c_p}{\hat q}^\subL
-\left(\frac{\hat h -\hat h_\subC^\subL}{\hat q}\right)^2\frac{\hat c_p^\subG}{\hat q}
\right)
\right],
\label{e:VL-pre}\\
&V^\subG-V_\eq^\subG=
|\phi|\frac{T_\subC}{2}
\frac{N}{\rho_\subC^\subG}\left[
\left(\frac{\hat h-\hat h_\subC^\subL}{\hat {q}}\right)^2
\alpha_\subC^\subG
\right.\nonumber\\
&~~~~~\left.
+
\left(
\left(\frac{\hat h_\subC^\subG-\hat h}{\hat q}\right)^2\frac{\hat c_p^\subL}{\hat q}
-\left(\frac{\hat h -\hat h_\subC^\subL}{\hat q}\right)^2\frac{\hat c_p^\subG}{\hat q}
\right)
\right],
\label{e:VG-pre}
\end{align}
where
 $\alpha_\subC^\subL$ and $\alpha_\subC^\subG$ are thermal expansion coefficients at constant pressure  for saturated liquid and gas.
  The scaled width of the liquid region, $x\equiv X/L_x$, changes as
\begin{align}
&x-x_\eq=
-|\phi|
\frac{T_\subC}{2}x_\eq (1-x_\eq)
\left[
\frac{\hat h-\hat h_\subC^\subL}{\hat q}\alpha_\subC^\subG
+
\frac{\hat h_\subC^\subG-\hat h}{\hat q}\alpha_\subC^\subL
\right.
\nonumber\\
&~~\left.
+\frac{\hat h_\subC^\subG-\hat h}{\hat h-\hat h_\subC^\subL}\frac{\hat c_p^\subL}{\hat q}
-\frac{\hat h-\hat h_\subC^\subL}{\hat h_\subC^\subG-\hat h}\frac{\hat c_p^\subG}{\hat q}
\right].
\label{e:Xneq-pre}
\end{align}
Using \eqref{e:xeq}, \eqref{e:Xneq-pre} is reformulated as
\begin{align}
&x-x_\eq=
-|\phi|\frac{T_\subC}{2}x_\eq (1-x_\eq)
\frac{\alpha_\subC^\subG\rho_\subC^\subG (1-x_\eq)+\alpha_\subC^\subL\rho_\subC^\subL x_\eq}{\rho_\subC^\subG (1-x_\eq)+\rho_\subC^\subL x_\eq}
\nonumber\\
&
-|\phi|\frac{T_\subC}{2}
\left[
\frac{\hat c_p^\subL}{\hat q}\frac{\rho_\subC^\subL}{\rho_\subC^\subG}x_\eq^2
-\frac{\hat c_p^\subG}{\hat q}\frac{\rho_\subC^\subG}{\rho_\subC^\subL}(1-x_\eq)^2
\right].
\label{e:Xneq-x}
\end{align}
When the liquid and gas satisfy $\rho_\subC^\subL\gg\rho_\subC^\subG$ and $\alpha_\subC^\subL\ll 1$, which are usually satisfied when $\Tc$ is much lower than the critical temperature, 
formula \eqref{e:Xneq-x} can be simplified as
\begin{align}
x-x_\eq
\simeq-|\phi|\frac{T_\subC}{2}\frac{\hat c_p^\subL}{\hat q}\frac{\rho_\subC^\subL}{\rho_\subC^\subG}x_\eq^2.
\label{e:deltax-u0}
\end{align}
Because $\rho_\subC^\subL\gg\rho_\subC^\subG$, the interface may shift drastically with a violent increase in volume when imposing a slight heat current while maintaining the amount of enthalpy.

\subsection{Steady-state expressions as functions of $(\bT,p,N,\phi)$}
\label{s:ss-TPN}

Owing to the thermodynamic equivalence, the above estimates can be translated to functions of $(\bT,p,N,\phi)$
once specifying the global temperature as $\bT(H,p,N,\phi)$ and its inverse function as $H(\bT,p,N,\phi)$.

As derived in Appendix \ref{s:bTs}, the global temperature for a given $(H,p,N,\phi)$ is written as
\begin{align}
\bT=\Tc+|\phi|\frac{T_\subC}{\hat q}\left(\hat h -\frac{\hat h_\subC^\subL+\hat h_\subC^\subG}{2}\right),
\label{e:bT-pre}
\end{align}
which has two limiting cases of $\bT\rightarrow T_\subC(1-|\phi|/2)$ for $\hat h\rightarrow \hat h_\subC^\subL$ and $\bT\rightarrow T_\subC(1+|\phi|/2)$ for $\hat h\rightarrow \hat h_\subC^\subG$.
The former limit indicates that $T_\subC(1-|\phi|)<T(x)<T_\subC$, where the system is almost completely occupied by liquid with $x\rightarrow 1$. 
The latter limit indicates that $T_\subC<T(x)<T_\subC(1+|\phi|)$ (i.e., $x\rightarrow 0$), where the system is occupied by gas.

Solving \eqref{e:bT-pre}, we obtain $\hat h$ as
\begin{align}
\hat h=\frac{\hat h_\subC^\subL+\hat h_\subC^\subG}{2}+\frac{\hat q}{|\phi|}\frac{\bT-T_\subC}{T_\subC}+O(\ep),
\label{e:h-bT}
\end{align}
which is a function of $(\bT,p,\phi)$. 
Note that the relation \eqref{e:h-bT} does not provide the value of $\hat h$ out of equilibrium owing to the error term of $O(\ep)$.
However, compared with the equilibrium liquid--gas coexistence at $T_\subC$ at which $\hat h$ can take an arbitrary value in $\hat h_\subC^\subL<\hat h<\hat h_\subC^\subG$, \eqref{e:h-bT} specifies a value of $\hat h$ uniquely in the equilibrium limit $|\phi|\rightarrow 0$ once we set $\bT$ as $\bT-T_\subC=r |\phi| T_\subC$ with a constant $r$; i.e., $\hat h$ is expressed as a function of $(p,r)$. For instance, 
for the system with $\rho_\subC^\subL\gg \rho_\subC^\subG$, \eqref{e:xeq} can be rewritten as
\begin{align}
x_\eq=\frac{\rho_\subC^\subG}{\rho_\subC^\subL}~\lim_{\phi\rightarrow 0}\frac{|\phi|T_\subC-2(\bT-T_\subC)}{|\phi|T_\subC+2(\bT-T_\subC)}
=\frac{\rho_\subC^\subG}{\rho_\subC^\subL}\frac{1-2r}{1+2r}.
\end{align}
From \eqref{e:deltax-u0}, 
we estimate the deviation of the interface as slight as $x-x_\eq \sim |\phi|\rho_\subC^\subG/\rho_\subC^\subL$
except for $\bT\rightarrow T_\subC(1-|\phi|/2)$ as $\hat h\rightarrow \hat h_\subC^\subL$.
There, the system is almost occupied by liquid  as $x_\eq\rightarrow 1$,
and 
the liquid--gas interface may jump with a sudden increase in the volume 
once a temperature difference between the two heat baths is imposed.

\section{Concluding Remarks}

\begin{figure}[bt]
\begin{center}
\includegraphics[scale=0.29]{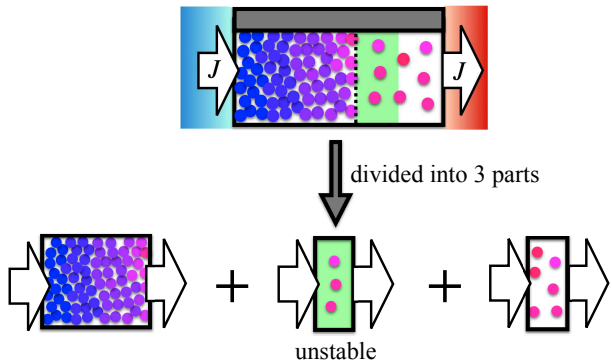}
\caption{Schematic explanation for the violation of additivity in heat conduction. If there is a region occupied by super-cooled gas, the local states are not kept in the separation of the system. }
\label{fig:nonadd}
\end{center}
\end{figure}

We derived a unique extension of the entropy function to heat conduction.
The extension works as a variational function whose maximum gives the  steady state and it maintains the fundamental thermodynamic relation by adopting the global temperature as the temperature of the whole system.
The different fundamental thermodynamic relations are connected by the Legendre transformation. 
See Table \ref{t:summary} for the overall view of the thermodynamic structure of the global thermodynamics.
The steady state is equivalent between the energy-conserving heat conduction system imposing a constant heat current and the standard heat conduction system in contact with two heat baths of different temperature,
which suggests the robustness of the steady state and the firm structure of the global thermodynamics.

The extended entropy is not an additive quantity but comprises the spatial integration of local entropy density and an extra term associated with phase coexistence.
Recalling that the additivity of extensive quantities is intrinsic in equilibrium thermodynamics,
such non-additivity is a manifestation of nonequilibrium.

Physically, the non-additivity may be related to the existence of the metastable state,
which appears in a macroscopic region adjacent to the interface.
As a thought experiment, we consider the separation of the system into three parts as shown in Fig.~\ref{fig:nonadd}, 
where the same heat current as for the original system is imposed on the three parts.
Immediately after the separation, the middle part is occupied by super-cooled gas, whose local temperature is $\theta<T(x)<T_\subC(p)$.
Once the middle part is separated from the other parts, 
this super-cooled gas is no longer in a steady state, and some portion of the middle part may transition into liquid. 
In this way, we expect that the heat conduction state with a liquid--gas interface is not additive once super-cooled gas is stabilized by the heat current.

We are able to transform the non-additive expression  \eqref{e:Sth-neq} into an additive form
by considering an isentropic surface, $S(H,p,N,\phi)=S(H_{\phi},p,N)$, which connects  a nonequilibrium state $(H,p,N,\phi)$ with an equilibrium state $(H_{\phi}, p, N)$.
Here, $H_\phi=H+T_\subC(p)\phi\Psi$ for $J\neq 0$.
The equilibrium entropy is the sum of that for liquid and that for gas. We thus write
\begin{align}
S(H,p,N,\phi)
=S(H_{\phi}^\subL,p,N_{\phi}^\subL)+S(H_{\phi}^\subG,p,N_{\phi}^\subG)
\label{e:S*}
\end{align}
with $H_{\phi}^\subL+H_{\phi}^\subG=H_\phi$ and $N_{\phi}^\subL+N_{\phi}^\subG=N$.
$H_\phi$, $H_{\phi}^\subLG$, and $N_{\phi}^\subLG$ are determined in Appendix \ref{s:additiveS}.
Although \eqref{e:S*} may look additive, we note that $H_{\phi}^\subL+H_{\phi}^\subG \neq H$ for $J\neq 0$,
and therefore, $S$ and $H$ cannot be simultaneously additive quantities no matter what transformation is performed.
Non-additivity is essential for the coexistence of liquid and gas under heat conduction.

When the transition temperature $T_\subC(p)$ is close to the critical temperature, the metastable region may disappear. 
Equation \eqref{e:theta-J} suggests that the interface temperature $\theta$ does not deviate from $T_\subC(p)$ because both thermal conductivity and density become equal in liquid and gas.
The extensive variable $\Psi$ vanishes because it is proportional to the latent heat $\hat q$.
Moreover, our initial assumption of the negligible thickness of the interface may not be valid at the critical point because fluctuations grow up and
the width of the liquid--gas interface may become macroscopic. 

We proposed in this paper a scheme with which we construct the variational function $\mathscr S$ using the correspondence with thermodynamic entropy $S$. For equilibrium systems, the variational function of thermodynamics is closely related to the probability density of unconstrained thermodynamic variables in fluctuation theory \cite{Callen}. 
This idea may apply to systems out of equilibrium.
We thus naturally conjecture that, in the setup we study, the probability density of the unconstrained thermodynamic variables  is also expressed by the variational function $\mathscr S$. 
To confirm the validity of the conjecture, we will study the potential for the fluctuation of thermodynamic variables by analyzing mesoscopic stochastic systems.

Apart from the maximum entropy principle,
there are several variational principles aimed at explaining nonequilibrium phenomena. In the linear response regime from equilibrium, the principle of minimum entropy production has been formulated \cite{MEP}. The principle states that the stationary distribution minimizes the entropy production rate \cite{MEP-measure}, which is well founded from the dynamical fluctuation theory based on statistical mechanics \cite{Maes-LD}. Moreover, as an attempt to apply the variational principle to various nonequilibrium phenomena, the principle of the maximum entropy production rate has also been studied \cite{MMEP}. However, these variational principles are not yet connected to the maximum entropy principle in thermodynamics. The extended maximum entropy principle proposed in this paper describes phenomena in the linear response regime, and the connection with the principle of minimum entropy production should thus be studied in the future. Furthermore, the concept of using global quantities  may be applied to a wide class of phenomena far from equilibrium. It may then be interesting to find an interface between our theory and variational principles far from equilibrium.

Last but not least, we remark that several important concepts of
nonequilibrium thermodynamics have not yet been studied. For example,
{\it excess heat} is the key concept for an extension of the Clausius
formula \cite{Landauer, Oono-Paniconi, Hatano-Sasa, Ruelle,KNST-SST,KNST-SST-Twist,Maes-SST, Jona-Lasinio-SST}.
The basic idea is that the heat in the time evolution
is decomposed into a house-keeping part and an excess part, where the former is identified as the part necessary for maintaining
the nonequilibrium steady state. However, there are several versions
of the decomposition, and it has not been established which decomposition is
most useful. We constructed an extended entropy for  the phase
coexistence in heat conduction, and we may therefore ask what is a measurable heat $Q$
defined by an extended Clausius equality $Q = \tilde T \Delta S$
with the entropy and the global temperature formulated in this paper.
If such a quantity $Q$ is explored in the present framework, 
the isentropic surface defined by $S(H,p,N,\phi)=S(H_{\phi},p,N)$ provides
{\it quasi-static adiabatic processes}. Generalizing the argument, we may
further consider general {\it adiabatic processes} in which the entropy never
decreases. 
There, $-T_\subC(p)\phi\Psi$, which corresponds to $H-H_\phi$, may be regarded as the minimum work that induces steady heat current.
It is a challenging problem to operationally identify adiabatic
conditions for heat conduction systems.

{\em Acknowledgment.---}  
The present study was supported by JSPS KAKENHI (Nos. 17H01148, 19K03647, 19H05795, 20K20425).

\appendix

\section{Preliminaries}

\subsection{Global thermodynamics for liquid and gas in heat conduction}
\label{s:GTD-LG}

In this section, we briefly review  global thermodynamics that holds in each region of liquid and gas. 
As derived in \cite{global-JSP}, liquid or gas in the linear response regime
of heat conduction is thermodynamically equivalent to
a corresponding equilibrium system.
This is because the profile of any thermodynamic density is approximately
linear and therefore all global thermodynamic quantities defined averaged
over the whole system are identified by the trapezoidal rule applied to
the integration. 

Below, we assume $J <0$ without loss of generality. 
Similarly to \eqref{e:bT},  global temperatures of liquid and gas are defined by
\begin{align}
\bT^\subL=\frac{\intL \rho(x)T(x)}{\intL \rho(x)},\quad
\bT^\subG=\frac{\intG \rho(x)T(x)}{\intG \rho(x)},
\label{e:TLTG}
\end{align}
and corresponding inverse temperatures by
\begin{align}
\bbeta^\subL=\frac{\intL \rho(x)\beta(x)}{\intL \rho(x)},\quad
\bbeta^\subG=\frac{\intG \rho(x)\beta(x)}{\intG \rho(x)},
\label{e:BLBG}
\end{align}
where $\beta(x)=1/T(x)$. We derive
\begin{align}
\bbeta^\subL=\frac{1}{\bT^\subL}+O(\ep^2), \quad
\bbeta^\subG=\frac{1}{\bT^\subG}+O(\ep^2),
\label{e:bLG-TLG}
\end{align}
which are the same as the relations at equilibrium.
Then, defining
\begin{align}
&\alpha^\subL=-\bbeta^\subL\mu^\subL(\bT^\subL, p)+O(\ep^2),\label{e:alphaL}\\
&\alpha^\subG=-\bbeta^\subG\mu^\subG(\bT^\subG, p)+O(\ep^2)\label{e:alphaG}
\end{align}
with equilibrium chemical potential $\mu^\subL(T,p)$ and $\mu^\subG(T,p)$
of liquid and gas, 
we have the fundamental relation of thermodynamics in each region as
\begin{align}
&dS^\subL={\bbeta^\subL}dH^\subL-\bbeta^\subL{V^\subL}dp+\alpha^\subL dN^\subL,\label{e:dS-L}\\
&dS^\subG={\bbeta^\subG}dH^\subG-{\bbeta^\subG}{V^\subG}dp+\alpha^\subG dN^\subG,\label{e:dS-G}
\end{align}
where $S^\subL$ and $S^\subG$ are the spatial integration of entropy density in \eqref{e:S-LG}.
These relations indicate
\begin{align}
S^\subL=S(H^\subL,p,N^\subL),\quad 
S^\subG=S(H^\subG,p,N^\subG).
\end{align}
That is, $S^\subL$ and $S^\subG$ are expressed by 
the equilibrium entropy function $S(H,p,N)$.
We emphasize that both liquid and gas are out of equilibrium
with conducting heat.

\subsection{Functions in liquid--gas coexistence}

Let $\cal H^\subl$ and $\cal N^\subl$ be the variational variables
describing  the enthalpy and  the particle number
in the left region of $0<x<X$.
In the argument below, six variable function
$A({\cal H}^\subl,{\cal N}^\subl;H,p,N,\phi)$ and four variable function
$A_*(H,p,N,\phi)$ appear. 
More precisely, for any  six variable function
$A({\cal H}^\subl,{\cal N}^\subl;H,p,N,\phi)$,  $A_*$ denotes the
four variable function defined by
\begin{align}
&A_*(H,p,N,\phi)=
\nonumber\\
&~~~
A(H^\subl(H,p,N,\phi),N^\subl(H,p,N,\phi);H,p,N,\phi),
\label{e:A-HpN}
\end{align}
where $H^\subl(H,p,N,\phi)$ and $N^\subl(H,p,N,\phi)$ are the steady-state values
when $(H,p,N,\phi)$ are given. Here, we assume that $\phi$ is also determined uniquely for a given $J$.
The relation \eqref{e:S*def} is simply  expressed by
$S={\mathscr S}_*$.
Obviously, $H^\subL={\cal H}^\subL_*$ and $N^\subL={\cal N}^\subL_*$

Although we do not specify the function type
once they are defined, the function types are fixed  throughout
this section, which is not the standard convention in thermodynamics.
Thus, for example, a partial derivative is simply expressed as
\begin{equation}
\pder{A}{H}.  
\end{equation}
Accordingly, the partial  derivative at the steady state is defined. For instance, the partial derivative with respect to $\cal H^\subl$  is written as
\begin{align}
\pderf{A}{\cal H^\subl}{*}=
\left. \pderf{A}{\cal H^\subl}{{\cal N}^\subl,H,p,N,\phi}\right|_{({\cal H^\subl}, {\cal N^\subl})=(H^\subl,N^\subl)}.
\label{e:dA*}
\end{align}

\subsection{Basic quantities for the liquid--gas coexistence}

Thermodynamic relations in each liquid or gas hold for
given ${\cal H}^\subl$ and ${\cal N}^\subl$. 
Therefore,  for the thermodynamic  states, $({\cal H}^\subl,p,{\cal N}^\subl)$ and $({\cal H}^\subr,p,{\cal N}^\subr)$ with ${\cal H}^\subr=H-{\cal H}^\subl$ and ${\cal N}^\subr=N-{\cal N}^\subl$, $\bbeta^\subl$ and $\bbeta^\subr$ can be defined from the fundamental thermodynamic relations \eqref{e:dS-L} and \eqref{e:dS-G} as
\begin{align}
\bbeta^\subl=\pderf{S^\subl}{\cal H^\subl}{p,{\cal N^\subl}}, \qquad
\bbeta^\subr=\pderf{S^\subr}{\cal H^\subr}{p,{\cal N^\subr}}.
\end{align}
We then introduce the global inverse temperature $\bbeta=\bbeta({\cal H}^\subl,{\cal N}^\subl;H,p,N)$ in the liquid--gas coexistence as
\begin{align}
\bbeta=\frac{{\cal N}^\subl}{N}\bbeta^\subl+\frac{{\cal N}^\subr}{N}\bbeta^\subr,
\label{e:bbeta-NLNG}
\end{align}
which is consistent with
\begin{align}
\bbeta=\frac{\intx \rho(x)\beta(x)}{\intx \rho(x)}.
\label{e:bbeta}
\end{align}
Developing a similar argument for $\alpha=-\mu/T$,
we have
\begin{align}
\alpha^\subl=\pderf{S^\subl}{\cal N^\subl}{{\cal H^\subl},p}, \quad
\alpha^\subr=\pderf{S^\subr}{\cal N^\subr}{{\cal N^\subr},p},
\end{align}
from which we define the global quantity in the liquid--gas coexistence
$\balpha=\balpha({\cal H}^\subl, {\cal N}^\subl;H,p,N,\phi)$ as
\begin{align}
\balpha=\frac{{\cal N}^\subl}{N}\alpha^\subl+\frac{{\cal N}^\subr}{N}\alpha^\subr
\label{e:balpha-NLNG}
\end{align}
consistently with
 \begin{align}
\balpha=\frac{\intx \rho(x)\alpha(x)}{\intx \rho(x)}.
\label{e:balpha}
\end{align}

In the steady state, we expect that the temperature profile $T(x)$ is continuous in $0<x<L_x$ including  the liquid--gas interface.
This is because there is no particle current to produce latent heat at the interface.
Since the temperature profile satisfies $T(x)=T_\subC(p)+O(\ep)$, we have $\bbeta_*=\beta_\subC(p)+O(\ep)$.
Let $\theta$ be  the temperature at the interface. 
The global temperatures of left and right regions are estimated as
\begin{align}
\bT^\subl_*=\frac{T_1+\theta}{2}+O(\ep^2),\quad
\bT^\subr_*=\frac{T_2+\theta}{2}+O(\ep^2)
\label{e:bT-LG}
\end{align}
by applying the trapezoidal rule to the integration \eqref{e:TLTG}.
These expressions result  in 
\begin{align}
\frac{\bT^\subr_*-\bT^\subl_*}{T_\subC(p)}=\frac{\ep}{2}+O(\ep^2),
\label{e:TG-TL}
\end{align}
which is equivalent to
\begin{align}
\frac{\bbeta^\subl_*-\bbeta^\subr_*}{\beta_\subC(p)}=\frac{\ep}{2}+O(\ep^2).
\label{e:bL-bG}
\end{align}
Together with \eqref{e:bbeta-NLNG}, this relation leads to 
\begin{align}
\bbeta^\subl_*=\bbeta_*\left(1+\frac{\ep}{2}\frac{N^\subr}{N}\right), \quad
\bbeta^\subr_*=\bbeta_*\left(1-\frac{\ep}{2}\frac{N^\subl}{N}\right).
\label{e:bLG-bb}
\end{align}

Multiplying the formula \eqref{e:bbeta-NLNG} and $\bT=({\cal N}^\subl \bT^\subl+{\cal N}^\subr \bT^\subr)/N$,
and applying \eqref{e:bLG-TLG}, we have
\begin{align}
\bbeta_*\bT_*=\left(\frac{N^\subl}{N}\right)^2+\left(\frac{N^\subr}{N}\right)^2+\frac{N^\subl N^\subr}{N^2}\left(\frac{\bbeta^\subl_*}{\bbeta^\subr_*}+\frac{\bbeta^\subr_*}{\bbeta^\subl_*}\right).
\label{e:bb-bT0}
\end{align}
Since \eqref{e:bLG-bb} leads to $\bbeta^\subl_*/\bbeta^\subr_*=1+\ep/2+O(\ep^2)$, we transform \eqref{e:bb-bT0} into
\begin{align}
\bbeta_*=\frac{1}{\bT_*}+O(\ep^2),
\label{e:bb-bT}
\end{align}
using $N^\subl+N^\subr=N$.
By a similar argument, we can show 
\begin{align}
\balpha_*=-\frac{\bmu_*}{\bT_*}+O(\ep^2).
\label{e:ba-bmu}
\end{align}
Thus, we can write the fundamental relation of thermodynamics
\eqref{e:1st-law} as
\begin{align}
dS=\bbeta_* dH-\bbeta_* V_* dp+\balpha_* dN +\Psi d\phi.
\label{e:1st-law-neq-x}
\end{align}

\section{Derivation of (\ref{e:phi}) and  (\ref{e:Psi})}
\label{sec:Psi-phi}

We first argue the concept of intensive and extensive variables in Appendix \ref{sec:int-extent}. 
Next, in Appendix \ref{det-psi-s}, we determine $\Psi$ from the variational equations and
fundamental thermodynamic relations. 
In Appendix \ref{s:singularity}, we formulate the singularity at $\ep=0$.
Finally, in Appendix \ref{sec:det-psi}, 
we determine $\psi$ and $\lambda$.
Appendix \ref{ap:interpretation} is devoted to discussion on the singularity.

\subsection{Intensive and extensive variables}
\label{sec:int-extent}

\begin{figure}[bt]
\begin{center}
\includegraphics[scale=0.35]{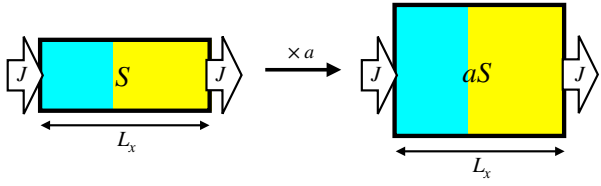}
\caption{Extensivity of the heat conduction state with a liquid--gas interface. $J$ is the amount of heat current per unit area. For a system with area $aL_y L_z$ in a cross section perpendicular to the $x$-axis, all profiles of local thermodynamic quantities are equal to those of the system with $L_y L_z$ if $J$ is the same.}
\label{fig:ExtAdd}
\end{center}
\end{figure}

Since heat conduction states are inhomogeneous along the horizontal direction parallel to the heat current,
the extensive property is not trivial for global thermodynamic variables.
On the other hand, thermodynamic states are homogeneous along the section perpendicular to the heat current, and therefore we can consider the extensive property for the size scaling procedure in this direction.
As schematically shown in  Fig. ~\ref{fig:ExtAdd},
we fix $L_x$ and $J$ in the argument below,
where $J$ is defined as the current per unit area. 
The total amount of heat current is proportional to $L_yL_z$.

We define an intensive variable $\phi$ indicating the degree of nonequilibrium,
which is proportional to $\ep$ 
and invariant  for the scaling transformation of $L_y L_z\rightarrow a L_y L_z$ with $H\rightarrow aH$ and $N\rightarrow aN$ in  Fig. ~\ref{fig:ExtAdd},
where $a$ is an arbitrary constant.
This invariance allows $\phi$ to depend on $\hat h=H/N$ and $p$.
Thus, we express $\phi$ as
\begin{align}
\phi=\frac{\ep}{\lambda(\hat h,p)}
\label{e:x}
\end{align}
without loss of generality, where $\lambda$ is an intensive function to be determined in the steady state.

As the conjugate variable to $\phi$, we set $\Psi$ as it is an extensive variable.
More precisely, $\Psi(H,p,N,\phi)$ is equal to the steady state value $\psi_*$ determined as \eqref{e:Psi*def} from $\psi({\cal H}^\subl,{\cal N}^\subl; H,p,N)$. 
For the scaling transformation in  Fig.~\ref{fig:ExtAdd}, the extensivity is expressed as
\begin{align}
&a \psi({\cal H}^\subl,{\cal N}^\subl; H,p,N)=\psi(a{\cal H}^\subl,a{\cal N}^\subl; aH,p,aN). \label{e:ext-Psi0}
\end{align}
Differentiating both sides in $a$ and setting $a=1$, 
we have the Euler relation,
\begin{align}
&  \psi=\pder{\psi}{{\cal H}^\subl} {{\cal H}^\subl}+\pder{\psi}{{\cal N}^\subl} {{\cal N}^\subl}+\pder{\psi}{H} {H}+\pder{\psi}{N} {N}.
\label{e:extensive*}
\end{align}

\subsection{Determination of $\psi$ and $\Psi$}
\label{det-psi-s}

Differentiating the variational function $\mathscr S$ in \eqref{e:Svar-neq}, we have
\begin{align}
&\pder{\mathscr S}{{\cal H}^\subl}=(\bbeta^\subl-\bbeta^\subr)+\phi\pder{\psi}{{\cal H}^\subl},\\
&\pder{\mathscr S}{{\cal N}^\subl}=(\alpha^\subl-\alpha^\subr)+\phi\pder{\psi}{{\cal N}^\subl},\\
&\pder{\mathscr S}{H}=\bbeta^\subr+\phi\pder{\psi}{H},\\
&\pder{\mathscr S}{p}=-\bbeta^\subl V^\subl-\bbeta^\subr V^\subr+\phi\pder{\psi}{p},\\
&\pder{\mathscr S}{N}=\alpha^\subr+\phi\pder{\psi}{N},\\
&\pder{\mathscr S}{\phi}=\psi. \label{e:S-phi}
\end{align}
Comparing these with \eqref{e:var-neq-HL}, \eqref{e:var-neq-NL}, \eqref{e:th-neq-H}, \eqref{e:th-neq-p}, \eqref{e:th-neq-N}, and \eqref{e:th-neq-phi},
we obtain five differential equations for $\psi$ at the steady state as
\begin{align}
&\phi\pderf{\psi}{{\cal H}^\subl}{*}=\bbeta^\subr_*-\bbeta^\subl_*,\label{e:Psi*-HL}\\
&\phi\pderf{\psi}{{\cal N}^\subl}{*}=\alpha^\subr_*-\alpha^\subl_*,\label{e:Psi*-NL}\\
&\phi\pderf{\psi}{H}{*}=\bbeta_*-\bbeta^\subr_*, \label{e:Psi*-H}\\
&\phi\pderf{\psi}{p}{*}=-\bbeta_* V_*+\bbeta^\subl_* V^\subl_*+\bbeta^\subr_* V^\subr_*, \label{e:Psi*-p}\\
&\phi\pderf{\psi}{N}{*}=\balpha_*-\alpha^\subr_* . \label{e:Psi*-N}
\end{align}
Substituting these five relations into \eqref{e:extensive*} at the steady state, we have
\begin{align}
\phi\Psi=-(\bbeta^\subl_*-\bbeta^\subr_*)\left(H^\subl-\frac{N^\subl}{N}H\right).
\end{align}
Using \eqref{e:bL-bG} and \eqref{e:x}, we further rewrite it as
\begin{align}
  \Psi=-\frac{\lambda(\hat h,p)}{2}\beta_\subC(p)
  \left(H^\subl-\frac{N^\subl}{N}H\right).
\label{e:Psi-star}  
\end{align}
The functional form of $\lambda(\hat h,p)$ still remains to be determined.

\subsection{Relation emerged from singularity}
\label{s:singularity}

\begin{figure}[b]
\begin{center}
\includegraphics[scale=0.43]{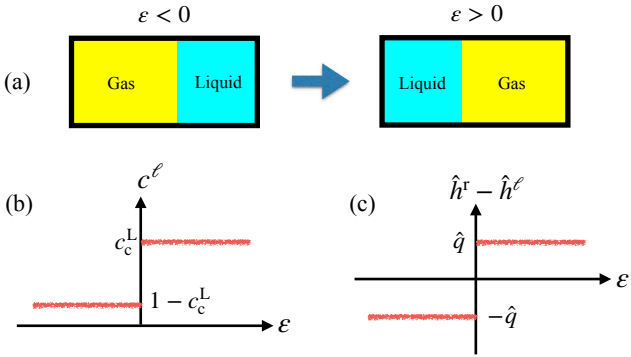}
\caption{(a) Transition of the steady state when $J$ is changed from $J>0$ ($\ep<0$) to $J<0$ ($\ep>0)$.
Correspondingly, the fraction of particles in the left region, $c^\subl=N^\subl/N$, behaves as a singular function of $\ep$ shown in (b). Similarly, $\hat h^\subr-\hat h^\subl$ shows discontinuous change as in (c).}
\label{fig:singular}
\end{center}
\end{figure}

Noting that the trivial relation $H=H^\subl+H^\subr$ with
the number fraction $c^\subl\equiv N^\subl/N$ and the enthalpy per particle
$\hat h^\sublr\equiv{H^\sublr}/{N^\sublr}$
 gives
\begin{align}
H^\subl-\frac{N^\subl}{N}H = -N
(\hat h^\subr-\hat h^\subl) (1-c^\subl)c^\subl,
\label{e:h-rL}
\end{align}
we rewrite \eqref{e:Psi-star} as
\begin{align}
{\Psi}={N}\frac{\lambda}{2}\beta_\subC(\hat h^\subr-\hat h^\subl)(1-c^\subl)c^\subl.
 \label{e:Psi*/N}
\end{align}
From the left--right symmetry, $\Psi$ should change its sign for $\ep \to -\ep$. 
Indeed, the expression in  \eqref{e:Psi*/N} explicitly shows this property
because $c^\subl \leftrightarrow c^\subr=1-c^\subl$ for $\ep \leftrightarrow -\ep$ and $\hat h^\subr-\hat h^\subl$ changes its sign at $\ep=0$ as shown in Fig. ~\ref{fig:singular}(c), where
$\hat q=\hat h_\subC^\subG-\hat h_\subC^\subL$ with
\begin{align}
&\lim_{\ep\rightarrow 0^+} \hat h^\subl(\phi)=\hat h_\subC^\subL,\quad
\lim_{\ep\rightarrow 0^-} \hat h^\subl(\phi)=\hat h_\subC^\subG,\\
&\lim_{\ep\rightarrow 0^+} \hat h^\subr(\phi)=\hat h_\subC^\subG,\quad
\lim_{\ep\rightarrow 0^-} \hat h^\subr(\phi)=\hat h_\subC^\subL.
\end{align}
The left side region of the interface is occupied by liquid for $\ep >0$ whereas by gas for $\ep <0$.
The position of the interface may jump at $\ep=0$ and the number fraction $c^\subl$ suddenly changes its value; i.e.,
\begin{align}
&\lim_{\ep\rightarrow 0^+} c^\subl(\phi)=c_\subC^\subL,\quad
\lim_{\ep\rightarrow 0^-} c^\subl(\phi)=1-c_\subC^\subL
\end{align}
with the equilibrium fraction of liquid $c_\subC^\subL$. See Fig. ~\ref{fig:singular}(b).

The singularity at $\ep=0$ is critical for the extension of entropy. 
To see it, we differentiate \eqref{e:Sth-neq} in $\phi$ with $H$, $p$,
and $N$ fixed and obtain
\begin{align}
\pder{S}{\phi}=
(\bbeta_*^\subl-\bbeta_*^\subr)\pder{H^\subl}{\phi}
+(\alpha^\subl_*-\alpha^\subr_*)\pder{N^\subl}{\phi}+\Psi+O(\ep)
\label{e:Sphi-tmp}
\end{align}
by noting $H^\subl=H^\subl(H,p,N,\phi)$ and $N^\subl=N^\subl(H,p,N,\phi)$.
Comparing \eqref{e:Sphi-tmp} with \eqref{e:th-neq-phi} and applying \eqref{e:bL-bG}, we obtain
\begin{align}
\frac{\ep}{2}\beta_\subC\pder{H^\subl}{\phi}
+(\alpha^\subl_*-\alpha^\subr_*)\pder{N^\subl}{\phi}=O(\ep).
\label{e:sin1}
\end{align}
At first glance, \eqref{e:sin1} is trivial because its left-hand side appears to be $O(\ep)$.
However, we should be careful to differentiate $H^\subl$ and $N^\subl$ in $\phi$ due to the singular behavior in Fig.~\ref{fig:singular}.
Using $\hat h^\subl$ and $c^\subl$, \eqref{e:sin1} is transformed into
\begin{align}
\left(\alpha^\subl_*-\alpha^\subr_*+\frac{\ep}{2}\beta_\subC\hat h\right)
\pder{c^\subl}{\phi}
=\frac{\ep}{2}\beta_\subC 
\pder{(\hat h^\subr-\hat h^\subl)c^\subl(1-c^\subl)}{\phi}.
\label{e:sin2}
\end{align}
We thus need to define the singular functions $c^\subl$ and $\hat h^\subl$.
We introduce a regularization parameter $\ep_1$ to
make $c^\subl(\phi)$ a smooth function in $\ep$:
\begin{align}
c^\subl(\phi)=\frac{1}{2}+\left(c_\subC^\subL-\frac{1}{2}\right)\tanh\left(\frac{\ep}{\ep_1}\right).\label{e:singularity-r}
\end{align}
Since $\hat h^\subr(\phi)-\hat h^\subl(\phi)$ is antisymmetric in $\ep$, 
we write
\begin{align}
\hat h^\subr(\phi)-\hat h^\subl(\phi)
=\hat q\tanh\left(\frac{\ep}{\ep_2}\right)
\label{e:singularity-h}
\end{align}
with another regularization parameter $\ep_2$.
Note that  $\ep_1$ and $\ep_2$ should be set to zero after the calculation.

Accepting these regularizations, we can calculate the derivatives with $\phi$ at $\ep=0$ as 
\begin{align}
\left.\pder{c^\subl}{\phi}\right|_{\ep=0}
=\lambda \frac{c_\subC^\subl-\frac{1}{2}}{\ep_1}, \quad
\left.\pder{(\hat h^\subr-\hat h^\subl)}{\phi}\right|_{\ep=0}
=\lambda \frac{\hat q}{\ep_2}.
\label{e:singularity-dif}
\end{align}
See Sec. \ref{ap:interpretation} for the interpretation of the regularization parameters.

Substituting the derivatives \eqref{e:singularity-dif} into \eqref{e:sin2}, we obtain
\begin{align}
\alpha^\subl_*-\alpha^\subr_*+\frac{\ep}{2}\beta_\subC\hat h
=\ep\frac{\ep_1}{\ep_2}
\beta_\subC\hat q\frac{c^\subl(1-c^\subl)}{2c^\subl-1},
\label{e:sin3}
\end{align}
where we ignored the contribution of $O(\ep^2)$.
Now, we consider the limit $\ep_1 \to 0$ and $\ep_2 \to 0$
  with $\ep_1/\ep_2$ fixed. The right-hand side of \eqref{e:sin3} diverges
  at $c^\subl=1/2$ if $\ep_1/\ep_2 \not =0$, while the left-hand side remains regular.
To avoid such inconsistency, the only possible case is that $\ep_1/\ep_2=0$, which
  leads to
\begin{align}
\alpha^\subl_*-\alpha^\subr_*+\frac{\ep}{2}\beta_\subC\hat h=0.
\label{e:alpha-balance}
\end{align}
We mention that \eqref{e:alpha-balance} 
 provides the balance of chemical potential in heat conduction.

\subsection{Determination of $\psi$ and $\lambda$}
\label{sec:det-psi}

Since the functional form of $\Psi$ is expressed as \eqref{e:Psi-star}, 
we set $\psi$ as 
\begin{align}
&\psi({\cal H}^\subl,{\cal N}^\subl;H,p,N)
=
-\frac{\cal B}{2}{\beta_\subC(p)}
\left({\cal H}^\subl-\frac{{\cal N}^\subl}{N}H\right),
\label{e:Psi-B}
\end{align}
where $\cal B$ is an intensive function
\begin{align}
{\cal B}={\cal B}\left(\frac{{\cal H}^\subl}{N},\frac{{\cal N}^\subl}{N},\frac{H}{N},p\right).
\end{align}
Note that $\cal B$ is independent of $\phi$ because $\psi$ does not depend on $\phi$.
To be consistent with \eqref{e:Psi-star}, 
we have
\begin{align}
{\cal B}_*=\lambda(\hat h,p).
\label{e:B*}
\end{align}

Differentiating \eqref{e:Psi-B}, we have
\begin{align}
&\pder{\psi}{{\cal H}^\subl}=-\frac{\cal B}{2}\beta_\subC(p)-\frac{\beta_c(p)}{2N}\left({\cal H}^\subl-\frac{{\cal N}^\subl}{N}H\right)\pder{\cal B}{({\cal H}^\subl/N)},\label{e:Psi-HL}\\
&\pder{\psi}{{\cal N}^\subl}=\frac{\cal B}{2}\beta_\subC(p)\frac{H}{N}-\frac{\beta_c(p)}{2N}\left({\cal H}^\subl-\frac{{\cal N}^\subl}{N}H\right)\pder{\cal B}{({\cal N}^\subl/N)},\label{e:Psi-NL}\\
&\pder{\psi}{H}=\frac{\cal B}{2}\beta_\subC(p)\frac{{\cal N}^\subl}{N}-\frac{\beta_c(p)}{2N}\left({\cal H}^\subl-\frac{{\cal N}^\subl}{N}H\right)\pder{\cal B}{(H/N)},\label{e:Psi-H}\\
&\pder{\psi}{p}=-\frac{\cal B}{2}\frac{d\beta_\subC}{dp}\left({\cal H}^\subl-\frac{{\cal N}^\subl}{N}H\right)-\frac{\beta_c(p)}{2}\left({\cal H}^\subl-\frac{{\cal N}^\subl}{N}H\right)\pder{\cal B}{p}.\label{e:Psi-p}
\end{align}
Comparing \eqref{e:Psi*-HL}, \eqref{e:Psi*-NL}, and \eqref{e:Psi*-H}
with \eqref{e:Psi-HL}, \eqref{e:Psi-NL}, and \eqref{e:Psi-H}, respectively,
we obtain
\begin{align}
&\pderf{\cal B}{({\cal H}^\subl/N)}{*}=0, \label{e:B-HL}\\
&\pderf{\cal B}{(H/N)}{*}=0, \label{e:B-H}\\
&\pderf{\cal B}{({\cal N}^\subl/N)}{*}=0,\label{e:B-NL}
\end{align}
where we have used
\eqref{e:bbeta-NLNG}, \eqref{e:balpha-NLNG}, \eqref{e:bL-bG}, \eqref{e:bLG-bb}, 
\eqref{e:alpha-balance}, and \eqref{e:B*}.
Meanwhile, \eqref{e:Psi*-p} and \eqref{e:Psi-p} bring a relation
\begin{align}
\pderf{\cal B}{p}{*}=-{\cal B_*}\left(\frac{1}{\beta_\subC}\frac{d\beta_\subC}{dp}+\frac{N^\subr V^\subl_*-N^\subl V^\subr_*}{N^\subr H^\subl-N^\subl H^\subr}\right).
\label{e:B-p-pre}
\end{align}
The Clausius--Clapeyron relation
\begin{align}
\frac{d\Tc}{dp}=
T_\subC \frac{\hat v^\subl-\hat v^\subr}{\hat h^\subl-\hat h^\subr}
\label{e:Clausius-Clapeyron}
\end{align}
leads to the simplification of \eqref{e:B-p-pre}.
Indeed, noting $N^\subr V_*^\subl-N^\subl V_*^\subr=(\hat v^\subl-\hat v^\subr)N^\subl N^\subr$ with specific volumes $\hat v^\sublr$ and $N^\subr H^\subl-N^\subl H^\subr=(\hat h^\subl-\hat h^\subr)N^\subl N^\subr$,
we obtain
\begin{align}
\pderf{\cal B}{p}{*}=0.
\label{e:B-p}
\end{align}
Summarizing \eqref{e:B-HL}, \eqref{e:B-H}, \eqref{e:B-NL}, and \eqref{e:B-p},
we find that $\cal B$ takes
the form
\begin{align}
\lambda_0+\!\!
\sum_{k+l\ge 2} \! a_{kl}  ({\cal H}^\subl \! -H^\subl(H,p,N,\phi))^k
({\cal N}^\subl \! -N^\subl(H,p,N,\phi) )^l
\end{align}
with a numerical constant $\lambda_0$, where 
$a_{kl}$ are functions of
$({\cal H}^\subl, {\cal N}^\subl,H,p,N)$. However, since $\cal B$ does not explicitly
depend on $\phi$, $H^\subl(H,p,N,\phi)$ should not appear in $\cal B$. This means
that $a_{kl}=0$, and  we conclude
\begin{align}
{\cal B}=\lambda_0.
\end{align}
Thus,  \eqref{e:Psi-B} is written as
\begin{align}
\psi=-\frac{\lambda_0}{2}\beta_\subC(p)\left({\cal H}^\subl-\frac{{\cal N}^\subl}{N}H\right).
\label{e:Psi-end}
\end{align}
Recalling  \eqref{e:Sth-neq}, \eqref{e:Svar-neq}, and the fundamental relation of thermodynamics \eqref{e:1st-law}, the value of the numerical constant $\lambda_0$ does not affect experimental observations at all. Thus, we set $\lambda_0=1$ without loss of generality.
We then reach \eqref{e:phi} and \eqref{e:Psi}.

\subsection{Interpretation of the regularization parameters}
\label{ap:interpretation}

Mathematically,  the regularization parameters $\ep_1$ and
$\ep_2$ are introduced to set  the discontinuous
  functions in Figs.~\ref{fig:singular}(b) and (c)
  to be a limit of continuous functions \eqref{e:singularity-r} and \eqref{e:singularity-h}.
 We here provide their interpretation. 	
 As demonstrated in Fig.~\ref{fig:symmetry}, two configurations, $(\subl,\subr)=(\subL,\subG)$ and 
  $(\subl,\subr)=(\subG,\subL)$, are equivalent at $\ep=0$ due to the left--right symmetry of the system.
The symmetry is violated by imposing heat current:
  In $\ep >0$, the probability of observing the configuration
  with $(\subl,\subr)=(\subL,\subG)$  becomes higher than that with
  $(\subl,\subr)=(\subG,\subL)$ because the temperature of the left region
  is lower than the right region. The probability approaches unity
  beyond some magnitude of $\ep$, which corresponds to the regularization parameters.	
For $\ep$ beyond the regularization parameters, we observe the  steady states as shown in Fig.~\ref{fig:singular}(a).

Recalling that the enthalpy and particle number are both conserved quantities,
their spatial distributions are not necessarily functionally  dependent. 
This indicates that $\ep_1$ can be intrinsically independent of $\ep_2$.
The two regularization parameters should satisfy $\ep_1/\ep_2 \to 0$ in the limit $\ep_1 \to 0$ and  $\ep_2 \to 0$ for the  present framework to be consistent as discussed in  Appendix \ref{s:singularity}. 
Although we do not have a physical interpretation of this result, we expect
that such regularization parameters may naturally appear in the analysis
of statistical mechanics of the phase separation in heat conduction. 
Performing molecular dynamics simulations of finite-size systems,
we observe the obvious separation of the liquid from gas when $|J|$ is not too much small.
The separation becomes subtle for sufficiently small but non-vanishing values of $|J|$.
For instance,
liquid and gas regions fluctuate and sometimes exchange, which leads to the smearing of the discontinuous change at $J=0$.
The crossover value of $|J|$ depends on the system size.
Characterizing such finite-size smearing effects may shed light on the properties of the regularization.

\section{Derivation of (\ref{e:LeGendre})}\label{s:LeGendre}

From local thermodynamic relation $h(x)-\mu(x)\rho(x)=T(x)s(x)$,
we have
\begin{align}
H-G=L_yL_z\intx T(x)s(x).
\label{e:ToLegedre}
\end{align}
Decompose the integral of the right-hand side as
\begin{align}
&\intx T(x)s(x)=
\nonumber\\
&~~~~
\bT \intx s(x)+\intx (T(x)-\bT)\rho(x)\hat s(x),
\label{e:decompo}
\end{align}
where $\hat s(x)=s(x)/\rho(x)$. 
$\hat s(x)$ is discontinuous at the interface; i.e., $\hat s(x)=\hat s^\subl(T(x),p)$ for $0<x<X$
and $\hat s(x)=\hat s^\subr(T(x),p)+O(\ep)$ for $X<x<L_x$.
We then estimate $\hat s^\sublr(T(x),p)=\hat s^\sublr_\subC+O(\ep)$,
where $\hat s^\sublr_\subC=\hat s^\sublr(\Tc(p),p)$.
Ignoring the contribution of $O(\ep^2)$, we transform  the last term of \eqref{e:decompo} as
\begin{align}
&\hat s^\subl_\subC \intL(T(x)-\bT)\rho(x)
+\hat s^\subr_\subC \intG(T(x)-\bT)\rho(x)
\nonumber\\
&=
\hat s^\subl_\subC (\bT^\subl-\bT)\frac{N^\subl}{L_yL_z}
+
\hat s^\subr_\subC (\bT^\subr-\bT)\frac{N^\subr}{L_yL_z},
\label{e:decompo2}
\end{align}
where the second line is obtained from the definition of the global temperature for each region.
From $N\bT = {N^\subl}\bT^\subl+ {N^\subr}\bT^\subr$, we also have
\begin{align}
 \bT^\subl-\bT = -\frac{N^\subr}{N}( \bT^\subr-\bT^\subl), 
\quad \bT^\subr-\bT = \frac{N^\subl}{N}( \bT^\subr-\bT^\subl).
\end{align}  
Substituting these two relations into \eqref{e:decompo2}, we express the right-hand side as
\begin{align}
\phi T_\subC(p)(\hat s^\subr_\subC-\hat s^\subl_\subC)\frac{N^\subl N^\subr}{2N}\frac{1}{L_yL_z},
\label{e:decompo3}
\end{align}
where we have used \eqref{e:bTL-st} and \eqref{e:bTG-st}.
Note that $\hat s^\subr_\subC-\hat s^\subl_\subC=\hat q\Tc$ when $(\subl,\subr)=(\subL,\subG)$
and $-\hat q\Tc$ when $(\subl,\subr)=(\subG,\subL)$. 
As $(\subl,\subr)=(\subL,\subG)$ for $\phi>0$ and $(\subG,\subL)$ for $\phi<0$, we further rewrite
\eqref{e:decompo3} as
\begin{align}
&|\phi|{\hat q(p)}\frac{N^\subL N^\subG}{2N}\frac{1}{L_yL_z}.
\label{e:D3}
\end{align}
Comparing \eqref{e:D3} with $\Psi$ in \eqref{e:Psi2}, we conclude
\begin{align}
\intx (T(x)-\bT)\rho(x)\hat s(x) =
\frac{T_\subC(p)\phi\Psi}{L_yL_z}.
\end{align}
Then, noting $\bT=T_\subC(p)+O(\ep)$,
we write \eqref{e:ToLegedre} as
\begin{align}
H-G=\bT \left(L_yL_x\intx s(x)+\phi\Psi\right),
\end{align}
which is \eqref{e:LeGendre}.

\section{Equivalence of \eqref{e:var-sol-NL} to \eqref{e:muL-muG}} \label{s:temp-relation}

According to \eqref{e:alphaL} and \eqref{e:alphaG}, we write
\begin{align}
\alpha^\subl=\alpha^\subl(\bbeta^\subl,p), \quad
\alpha^\subr=\alpha^\subr(\bbeta^\subr,p) .
\end{align}
We expand  these functions  around $\beta_\subC(p)$ as 
\begin{align}
&\alpha^\subl=\alpha^\subl(\beta_\subC(p),p)-(\bbeta^\subl-\beta_\subC(p))\hat h^\subl+O(\ep^2),\\
&\alpha^\subr=\alpha^\subr(\beta_\subC(p),p)-(\bbeta^\subr-\beta_\subC(p))\hat h^\subr+O(\ep^2),
\end{align}
where we have used $\left({\partial \alpha}/{\partial \beta}\right)_{p}=-\hat h$
resulted from the Gibbs-Duhem relation. Let us substitute  these equations
into a re-expressed form of \eqref{e:var-sol-NL}; i.e., into
\begin{align}
\alpha^\subl-\alpha^\subr+\phi\beta_\subC(p)\frac{H}{2N}=0.
\end{align}
Then, using $\alpha^\subl(\beta_\subC(p),p)=\alpha^\subr(\beta_\subC(p),p)$ and $H=\hat h^\subl N^\subl+\hat h^\subr N^\subr$, 
we obtain
\begin{align}
-(\bbeta^\subl-\beta_\subC(p))\hat h^\subl
&+(\bbeta^\subr-\beta_\subC(p))\hat h^\subr 
\nonumber \\
+&\frac{\phi\bbeta}{2}\left(\frac{N^\subl}{N}\hat h^\subl+\frac{N^\subr}{N}\hat h^\subr\right)=0,
\label{e:C5}
\end{align}
where $\phi\beta_\subC(p)$ is replaced  by $\phi\bbeta$ by ignoring the difference of $O(\ep^2)$. 
Substituting \eqref{e:bLG-bb} into $\bbeta^\subl$ and $\bbeta^\subr$,
we further transform \eqref{e:C5} into
\begin{align}
(\hat h^\subr-\hat h^\subl) \left[\bbeta\left(1+\frac{\phi}{2}\frac{N^\subr-N^\subl}{N}\right)-\beta_\subC(p)\right]=0.
\end{align}
Since $|\hat h^\subr-\hat h^\subl|$ at $\ep=0$ corresponds to the latent heat $\hat q $, $\hat h^\subr-\hat h^\subl$ is non-zero.
Thus, we conclude that the variational principle \eqref{e:var-sol-NL}
is written as 
\begin{align}
\bbeta=\beta_\subC(p)\left(1-\frac{\phi}{2}\frac{N^\subr-N^\subl}{N}\right),
\label{e:bb-st}
\end{align}
which is equal to
\begin{align}
\bT=T_\subC(p)\left(1+\frac{\phi}{2}\frac{N^\subr-N^\subl}{N}\right).
\label{e:bT-st}
\end{align}
Substituting \eqref{e:bb-st} into \eqref{e:bLG-bb}, we have
\begin{align}
&\bT^\subl=T_\subC(p)\left(1-\frac{\phi}{2}\frac{N^\subl}{N}\right),\label{e:bTL-st}\\
&\bT^\subr=T_\subC(p)\left(1+\frac{\phi}{2}\frac{N^\subr}{N}\right).\label{e:bTG-st}
\end{align}

We can repeat a similar procedure starting with \eqref{e:muL-muG}.
We expand the two chemical potentials $\mu^\subl=\mu^\subl(\bT^\subl,p)$ and $\mu^\subr=\mu^\subr(\bT^\subr,p)$ as 
\begin{align}
&\mu^\subl=\mu^\subl(T_\subC(p),p)-(\bT^\subl-T_\subC(p))\hat s^\subl+O(\ep^2),\\
&\mu^\subr=\mu^\subr(T_\subC(p),p)-(\bT^\subr-T_\subC(p))\hat s^\subr+O(\ep^2),
\end{align}
where we have used $ (\partial \mu/\partial T)_{p}=-\hat s$.
Here, $\bT^\subl$ and $\bT^\subr$ are estimated by \eqref{e:bLG-bb} with $\bT^\subl=1/\bbeta^\subl$ and $\bT^\subr=1/\bbeta^\subr$.
Substituting these into \eqref{e:muL-muG} and noting that $\mu^\subl(T_\subC(p),p)=\mu^\subr(T_\subC(p),p)$, 
we have
\begin{align}
(\hat s^\subr-\hat s^\subl)\left(\bT-T_\subC(p)-\frac{\phi\bT}{2}\frac{N^\subr-N^\subl}{N}\right)=0.
\end{align}
Since $\hat s^\subr\neq \hat s^\subl$, we obtain the same relation as \eqref{e:bT-st}. That is, the steady state is equivalent.

We add a comment.
Summing up the last two formulas and using \eqref{e:bT-LG}
and \eqref{e:bT-st}, we have
the temperature relation
\begin{align}
\bT-\frac{T_1+T_2}{2}=\theta-\Tc(p),
\end{align}
which was derived in \cite{global-JSP}.

\section{Additivity  of $G$ and $F$}\label{s:add-GF}

According to \eqref{e:bG-def} and \eqref{e:bF-def}, the free energies are expressed by additive forms,
\begin{align}
&G=G^\subL+G^\subG,
\label{e:Gth-neq}
\\
&F=F^\subL+F^\subG,
\label{e:Fth-neq}
\end{align}
where $G^\subL=L_yL_z\intL \mu^\subL(x)\rho(x)$, $G^\subG=L_yL_z\intG \mu^\subG(x)\rho(x)$,
$F^\subL=L_yL_z\intL f(x)$, and $F^\subG=L_yL_z\intG f(x)$.
As has been derived in \cite{global-JSP} and shortly reviewed in Appendix \ref{s:GTD-LG}, respective global thermodynamics hold in liquid and gas, and therefore
\begin{align}
&G^\subL=\mu^\subL(\bT^\subL,p)N^\subL,\quad
G^\subG=\mu^\subG(\bT^\subG,p)N^\subG,\\
&F^\subL=F(\bT^\subL,V^\subL, N^\subL), \quad
F^\subG=F(\bT^\subG,V^\subG, N^\subG)
\end{align}
with respective global temperatures
\begin{align}
\bT^\subL=\bT\left(1-\phi \frac{N^\subG}{2N}\right),\quad
\bT^\subG=\bT\left(1+\phi\frac{N^\subL}{2N}\right),
\label{e:TLTG-bT}
\end{align}
which are consistent with \eqref{e:var-TLTG}.

We notice $\bT^\subL\neq \bT^\subG\neq\bT$ for $\ep\neq 0$.
Respecting the global temperature $\bT$, we rewrite \eqref{e:Gth-neq} and \eqref{e:Fth-neq} as parallel expressions to the non-additive entropy \eqref{e:Sth-neq};
\begin{align}
&G=G^\subL(\bT,p,N^\subL)+G^\subG(\bT,p,N^\subG)-\phi \bT{\Psi},
\label{e:Gth-neq-nonad}
\\
&F=F(\bT,V^\subL,N^\subL)+F(\bT,V^\subG,N^\subG)-\phi\bT {\Psi}
\label{e:Fth-neq-nonad}
\end{align}
by expanding $G^\subLG$ and $F^\subLG$ around $\bT$ and ignoring the contribution of $O(\ep^2)$.
Here,  $G^\subL(\bT,p,N^\subL)=\mu^\subL(\bT,p)N^\subL$ and $G^\subG(\bT,p,N^\subG)=\mu^\subG(\bT,p)N^\subG$.
Despite the apparent consistency of \eqref{e:Gth-neq-nonad} and \eqref{e:Fth-neq-nonad} with \eqref{e:Sth-neq} for the non-additive entropy, we
do not adopt $\bT$ as the temperature to characterize liquid and gas
because 
the fundamental thermodynamic relations of liquid and gas are \eqref{e:dS-L} and \eqref{e:dS-G} with $\bT^\subLG$ as each global temperature. Indeed, these fundamental relations provide
\begin{align}
dS^\subLG\neq \frac{dH^\subLG}{\bT}-\frac{V^\subLG}{\bT}dp-\frac{\mu^\subLG(\bT,p)}{\bT}dN^\subLG,
\end{align}
which implies that $\bT$ does not work as global temperature
for liquid or gas. Nevertheless, we can show
\begin{align}
&S^\subL=\frac{H^\subL-G^\subL(\bT^\subL,p,N^\subL)}{\bT^\subL}
=\frac{H^\subL-G^\subL(\bT,p,N^\subL)}{\bT},
\\
&S^\subG=\frac{H^\subG-G^\subG(\bT^\subG,p,N^\subG)}{\bT^\subG}
=\frac{H^\subG-G^\subG(\bT,p,N^\subL)}{\bT}.
\end{align}
This suggests that $\tilde T$ plays some role of the temperature in each region.

\section{Derivations of relations in Sec. \ref{s:steady-state}}\label{s:steady-state-derive}

We derive various relations for the steady state in heat conduction reported in Sec. \ref{s:steady-state} with assuming $J\le 0$. 
We can change $\phi$ and $J$ into $|\phi|$ and $|J|$ for the obtained relations 
with taking care of $\phi J\le 0$, because these formulas are common in  $J\ge 0$.  
We then arrive at the formulas in  Sec. \ref{s:steady-state}.

\subsection{Preliminaries}
We introduce a ratio representing the particle distribution as
\begin{align}
r\equiv \frac{1}{2}-\frac{N^\subL}{N} ,
\label{e:rdef-S3}
\end{align}
by which we have
\begin{align}
\frac{N^\subL}{N}=-r+\frac{1}{2},\quad
\frac{N^\subG}{N}=r+\frac{1}{2}.
\label{e:r-NLNG}
\end{align}
We first consider equilibrium liquid--gas coexistence with $(H,p,N)$ fixed.
Let $H^\subLG_\eq$ and $N^\subLG_\eq$ be the enthalpy and the
particle number of liquid and gas at equilibrium. 
In the coexistence state at  given $p$, there are constraints
among $H_\eq^\subLG$ and $N_\eq^\subLG$ such that
\begin{align}
H_\eq^\subL=\hat h_\subC^\subL(p) N_\eq^\subL, \quad
H_\eq^\subG=\hat h_\subC^\subG(p) N_\eq^\subG.
\end{align}
Combining with $H=H_\eq^\subL+H_\eq^\subG$ and $N=N_\eq^\subL+N_\eq^\subG$,
we have
\begin{align}
r_\eq=\frac{\hat h -\hat h_\subM(\pex)}{\hat q(\pex)},
\label{e:req-S3}
\end{align}
where 
\begin{align}
\hat h_\subM(p)=\frac{\hat h_\subC^\subL(p)+\hat h_\subC^\subG(p)}{2}.
\label{e5}
\end{align}

\subsection{Derivation of \eqref{e:bT-pre}}
\label{s:bTs}

The formulas \eqref{e:bT-st}, \eqref{e:bTL-st}, and \eqref{e:bTG-st}
for the global temperatures $\bT$, $\bT^\subL$, and $\bT^\subG$
are rewritten with $r$ as
\begin{align}
&\bT=T_\subC(1+\phi r) +O(\ep^2),\label{e:bT-S3-0}\\
&\bT^\subL=\Tc\left(1+\frac{\phi}{2}\left(r-\frac{1}{2}\right)\right),\label{e:bTL-S3-0}\\
&\bT^\subG=\Tc\left(1+\frac{\phi}{2}\left(r+\frac{1}{2}\right)\right)\label{e:bTG-S3-0}.
\end{align}
By substituting \eqref{e:req-S3} with \eqref{e5} into these equations,
we obtain
\begin{align}
&\bT=\Tc\left(1+\phi\frac{\hat h -\hat h_\subM}{\hat q}\right),\label{e:bT-S3}\\
&\bT^\subL=\Tc\left(1-\frac{\phi}{2}\frac{\hat h_\subC^\subG -\hat h}{\hat q}\right),\label{e:bTL-S3}\\
&\bT^\subG=\Tc\left(1+\frac{\phi}{2}\frac{\hat h-\hat h_\subC^\subL}{\hat q}\right).\label{e:bTG-S3}
\end{align}
Here, \eqref{e:bT-S3} corresponds to \eqref{e:bT-pre}.

\subsection{Derivation of  \eqref{e:HL-pre}, \eqref{e:HG-pre} and
  \eqref{e:NL-pre}, \eqref{e:NG-pre} }

As demonstrated in Appendix \ref{s:GTD-LG}, heat conduction states for liquid and gas  in the linear response regime are mapped to equilibrium states.
We then express 
\begin{align}
H^\subL=\hat h^\subL(\bT^\subL,p) N^\subL.
\end{align}
Substituting \eqref{e:bTL-S3} into it and then expanding
$\hat h^\subL$ in $\ep$,
we obtain
\begin{align}
H^\subL=\left(\hat h_\subC^\subL-\phi\frac{T_\subC}{2}\hat c_p^\subL\frac{\hat h_\subC^\subG -\hat h}{\hat q}
\right)N^\subL,
\label{e:HL-S3}
\end{align}
which is \eqref{e:HL-pre}.
Similarly, we have
\begin{align}
H^\subG=\left(\hat h_\subC^\subG+\phi\frac{T_\subC}{2}\hat c_p^\subG\frac{\hat h-\hat h_\subC^\subL}{\hat q}
\right)N^\subG,
\label{e:HG-S3}
\end{align}
which is \eqref{e:HG-pre}.
Recalling that $H=H^\subL+H^\subG$ and
applying \eqref{e:r-NLNG} and \eqref{e:req-S3},
we have
\begin{align}
&\hat h =
  \left(\frac{1}{2}-r\right)\hat h_\subC^\subL
  +\left(\frac{1}{2}+r\right)\hat h_\subC^\subG
\nonumber\\
&
-\phi\frac{T_\subC}{2}
\left[
\left(\frac{\hat h_\subC^\subG-\hat h}{\hat q}\right)^2\hat c_p^\subL
-\left(\frac{\hat h -\hat h_\subC^\subL}{\hat q}\right)^2\hat c_p^\subG
\right].
\label{e:h-S3}
\end{align}
Solving \eqref{e:h-S3} in $r$,  we obtain
\begin{align}
r=
\frac{\hat h -\hat h_\subM}{\hat q}
+
\phi\frac{T_\subC}{2\hat q}
\left[
\left(\frac{\hat h_\subC^\subG-\hat h}{\hat q}\right)^2\hat c_p^\subL
-\left(\frac{\hat h -\hat h_\subC^\subL}{\hat q}\right)^2\hat c_p^\subG
\right].
\label{e:r-S3}
\end{align}
Thus, the value of $r$ is uniquely determined for given $\hat h$ and $p$.
Explicitly, the number of particles is written as
\begin{align}
&\frac{N^\subL}{N}=
\frac{\hat h_\subC^\subG-\hat h}{\hat q}
-
\phi\frac{T_\subC}{2\hat q}
\left[
\left(\frac{\hat h_\subC^\subG-\hat h}{\hat q}\right)^2\hat c_p^\subL
-\left(\frac{\hat h -\hat h_\subC^\subL}{\hat q}\right)^2\hat c_p^\subG
\right],
\label{e:NL-S3}
\\
&\frac{N^\subG}{N}=
\frac{\hat h-\hat h_\subC^\subL}{\hat q}
+
\phi\frac{T_\subC}{2\hat q}
\left[
\left(\frac{\hat h_\subC^\subG-\hat h}{\hat q}\right)^2\hat c_p^\subL
-\left(\frac{\hat h -\hat h_\subC^\subL}{\hat q}\right)^2\hat c_p^\subG
\right],
\label{e:NG-S3}
\end{align}
which correspond to \eqref{e:NL-pre} and \eqref{e:NG-pre}.

\subsection{Derivation of  \eqref{e:VL-pre} and \eqref{e:VG-pre}}

The volumes of liquid and gas are connected to the densities as
\begin{align}
V^\subLG=\frac{N^\subLG}{\rho^\subLG}=\frac{N^\subLG}{N}\frac{N}{\rho^\subLG}.
\label{e:Vdef}
\end{align}
By substituting \eqref{e:bTL-S3-0} and \eqref{e:bTG-S3-0} into
$\rho^\subL=\rho^\subL(\bT^\subL,p)$  and $\rho^\subG=\rho^\subG(\bT^\subG,p)$
and expanding them in $\ep$, we have
\begin{align}
&\rho^\subL=\rho_\subC^\subL
\left[
1-\phi\frac{T_\subC}{2}\left(r-\frac{1}{2}\right)\alpha_\subC^\subL
\right],\label{e:rhoL-neq}\\
&\rho^\subG=\rho_\subC^\subG
\left[
1-\phi\frac{T_\subC}{2}\left(r+\frac{1}{2}\right)\alpha_\subC^\subG
\right],\label{e:rhoG-neq}
\end{align}
where  $\alpha^\subL$ and $\alpha^\subG$ are thermal expansion
coefficients of liquid and gas at constant pressure, defined by
\begin{align}
\alpha^\subL=-\frac{1}{\rho^\subL}\pderf{\rho^\subL}{T}{p},\quad
\alpha^\subG=-\frac{1}{\rho^\subG}\pderf{\rho^\subG}{T}{p}.
\end{align}
$\alpha_\subC^\subL$ and $\alpha_\subC^\subG$ are the values at the saturation.
Thus, \eqref{e:Vdef} leads to
\begin{align}
&V^\subL=\frac{N}{\rho_\subC^\subL}\frac{N^\subL}{N}
\left[
1+\phi\frac{T_\subC}{2}\left(r-\frac{1}{2}\right)\alpha_\subC^\subL
\right],\\
&V^\subG=\frac{N}{\rho_\subC^\subG}\frac{N^\subG}{N}
\left[
1+\phi\frac{T_\subC}{2}\left(r+\frac{1}{2}\right)\alpha_\subC^\subG
\right].
\end{align}
Substituting \eqref{e:r-S3}, \eqref{e:NL-S3}, and \eqref{e:NG-S3}
into these, $V^\subL$ and $V^\subG$ for given $(\hat h, \pex,\phi)$
are written  as
\begin{align}
&V^\subL=\frac{N}{\rho_\subC^\subL}\left[
\frac{\hat h_\subC^\subG-\hat h}{\hat {q}}
-\phi\frac{T_\subC}{2}
\left(\frac{\hat h_\subC^\subG-\hat h}{\hat {q}}\right)^2
\alpha_\subC^\subL
\right.\nonumber\\
&~~~\left.
-\phi\frac{T_\subC}{2\hat q}
\left(
\left(\frac{\hat h_\subC^\subG-\hat h}{\hat q}\right)^2\hat c_p^\subL
-\left(\frac{\hat h -\hat h_\subC^\subL}{\hat q}\right)^2\hat c_p^\subG
\right)
\right],
\label{e:VL-S3}\\
&V^\subG=\frac{N}{\rho_\subC^\subG}\left[
\frac{\hat h-\hat h_\subC^\subL}{\hat {q}}
+\phi\frac{T_\subC}{2}
\left(\frac{\hat h-\hat h_\subC^\subL}{\hat {q}}\right)^2
\alpha_\subC^\subL
\right.\nonumber\\
&~~~\left.
+\phi\frac{T_\subC}{2\hat q}
\left(
\left(\frac{\hat h_\subC^\subG-\hat h}{\hat q}\right)^2\hat c_p^\subL
-\left(\frac{\hat h -\hat h_\subC^\subL}{\hat q}\right)^2\hat c_p^\subG
\right)
\right],
\label{e:VG-S3}
\end{align}
which are \eqref{e:VL-pre} and \eqref{e:VG-pre}.

\subsection{Derivation  of  \eqref{e:Xneq-pre}}

We define the normalized position of the liquid--gas interface as
\begin{align}
x\equiv\frac{X}{L_x}=\frac{V^\subL}{V^\subL+V^\subG}.
\label{e:x-VLVG}
\end{align}
From \eqref{e:VL-S3} and \eqref{e:VG-S3}, the volumes in equilibrium
are written as
\begin{align}
V_\eq^\subL=\frac{N}{\rho_\subC^\subL}\frac{\hat h_\subC^\subG-\hat h}{\hat q}, \quad
V_\eq^\subG=\frac{N}{\rho_\subC^\subG}\frac{\hat h-\hat h_\subC^\subL}{\hat q},
\label{e:VLVG-eq}
\end{align}
and the ratio of the volume in heat conduction deviates from the equilibrium ratio as
\begin{align}
\frac{V^\subG}{V^\subL}=\frac{V_\eq^\subG}{V_\eq^\subL}
\left(
1+\phi\frac{T_\subC}{2}A
\right)
,
\label{e:VG/VL}
\end{align}
where
\begin{align}
A&=
\alpha_\subC^\subL 
\frac{\hat h -\hat h_\subC^\subL}{\hat q}
+
\alpha_\subC^\subG 
\frac{\hat h_\subC^\subG-\hat h}{\hat q}
+
\frac{\hat c_p^\subL}{\hat q}
\frac{\hat h_\subC^\subG-\hat h}{\hat h -\hat h_\subC^\subL}
-
\frac{\hat c_p^\subG}{\hat q}
\frac{\hat h -\hat h_\subC^\subL}{\hat h_\subC^\subG-\hat h}
.
\end{align}
From \eqref{e:x-VLVG} and \eqref{e:VG/VL}, it is found that 
\begin{align}
x-x_\eq&=
-\phi\frac{T_\subC}{2}\left(1+\frac{V_\eq^\subG}{V_\eq^\subL}\right)^{-2}\frac{V_\eq^\subG}{V_\eq^\subL} A.
\end{align}
Here, the equilibrium position $x_\eq$ is calculated from \eqref{e:x-VLVG} and \eqref{e:VLVG-eq} as
\begin{align}
x_\eq=\frac{\rho_\subC^\subG(\hat h_\subC^\subG-\hat h)}{\rho_\subC^\subL(\hat h-\hat h_\subC^\subL)+\rho^\subG(\hat h_\subC^\subG-\hat h)}.
\label{e:x-eq}
\end{align}
Since $V_\eq^\subG/V_\eq^\subL$ is connected to $x_\eq$, we have
\begin{align}
&x-x_\eq=-\phi\frac{T_\subC}{2}x(1-x_\eq)A
\nonumber\\
&=
-\phi\frac{T_\subC}{2}x_\eq(1-x_\eq)
\times
\nonumber\\
&\left[
\alpha_\subC^\subL 
\frac{\hat h -\hat h_\subC^\subL}{\hat q}
+
\alpha_\subC^\subG 
\frac{\hat h_\subC^\subG-\hat h}{\hat q}
+
\frac{\hat c_p^\subL}{\hat q}
\frac{\hat h_\subC^\subG-\hat h}{\hat h -\hat h_\subC^\subL}
-
\frac{\hat c_p^\subG}{\hat q}
\frac{\hat h -\hat h_\subC^\subL}{\hat h_\subC^\subG-\hat h}
\right],
\end{align}
which corresponds to  \eqref{e:Xneq-pre}.

\subsection{Derivation of \eqref{e:phi-J}}

Since the steady heat current $J$ is homogeneous in the container, we have
\begin{align}
J=-\kappa_\subC^\subL\frac{\theta-T_1}{X}
=-\kappa_\subC^\subG\frac{T_2-\theta}{L_x-X},
\label{e:J-theta}
\end{align}
which leads to
\begin{align}
\theta=\frac{\kappa_\subC^\subL(L_x-X)T_1+\kappa_\subC^\subG X T_2}{\kappa_\subC^\subL(L_x-X)+\kappa_\subC^\subG X}.
\end{align}
Using this expression, \eqref{e:J-theta} is written as
\begin{align}
J=-\frac{\kappa_\subC^\subL\kappa_\subC^\subG}{\kappa_\subC^\subL(1-x_\eq)+\kappa_\subC^\subG x_\eq}\frac{T_2-T_1}{L_x}.
\label{e:J-xeq}
\end{align}
Substituting \eqref{e:x-eq} into \eqref{e:J-xeq}, we arrive at \eqref{e:phi-J}.

\subsection{Derivation of \eqref{e:theta-J}}\label{s:theta-J}

The uniformity of the heat current in $x$,
\begin{align}
J=-\kappa_\subC^\subL\frac{\theta-T_1}{X}=-\kappa_\subC^\subG\frac{T_2-\theta}{L_x-X},
\end{align}
leads to
\begin{align}
&\bT^\subL=\theta+J\frac{X}{2\kappa_\subC^\subL},
\label{e:TL-J}\\
&\phi=-\frac{J}{T_\subC(p)}\left(\frac{X}{\kappa_\subC^\subL}+\frac{L_x-X}{\kappa_\subC^\subG}\right),
\label{e:phi-J-app}
\end{align}
where $\bT^{\subL}=(T_1+\theta)/2$ and $\phi=(T_2-T_1)/T_\subC(p)$.
Substituting \eqref{e:TL-J} and \eqref{e:phi-J-app} into \eqref{e:bTL-st}, we obtain
\begin{align}
\theta=T_\subC(p)-\frac{J L_x}{2}\left(\frac{X}{L_x}\frac{N^\subG}{N}\frac{1}{\kappa_\subC^\subL}-\frac{L_x-X}{L_x}\frac{N^\subL}{N}\frac{1}{\kappa_\subC^\subG}\right).
\label{e:theta-J-1}
\end{align}
Since \eqref{e:Vdef} yields
\begin{align}
\frac{N^\subL}{N}=\frac{X}{L_x}\frac{\rho^\subL}{\bar \rho}, \quad
\frac{N^\subG}{N}=\frac{L_x-X}{L_x}\frac{\rho^\subG}{\bar \rho}, 
\end{align}
where $\bar \rho=N/V$, we write \eqref{e:theta-J-1} as
\begin{align}
\theta=T_\subC(p)-
J\left(\frac{\rho_\subC^\subG}{\bar\rho}\frac{1}{\kappa_\subC^\subL}-\frac{\rho_\subC^\subL}{\bar\rho}\frac{1}{\kappa_\subC^\subG}\right)\frac{X(L_x-X)}{2L_x}.
\label{e:theta-J-2}
\end{align}
Using
\begin{align}
\bar\rho=\frac{X}{L_x}\rho^\subL+\frac{L_x-X}{L_x}\rho^\subG,
\end{align}
we have
\begin{align}
\frac{\rho_\subC^\subG}{\bar\rho}\frac{1}{\kappa_\subC^\subL}-\frac{\rho_\subC^\subL}{\bar\rho}\frac{1}{\kappa_\subC^\subG}
&=
\frac{1}{\kappa_\subC^\subL}-\frac{1}{\kappa_\subC^\subG}
+\left(\frac{X}{\kappa_\subC^\subL L_x}+\frac{L_x-X}{\kappa_\subC^\subG L_x}\right)\frac{\rho_\subC^\subL-\rho_\subC^\subG}{\bar\rho}\\
&=
\frac{1}{\kappa_\subC^\subL}-\frac{1}{\kappa_\subC^\subG}
-\frac{\phi T_\subC}{J L_x}
\frac{\rho_\subC^\subL-\rho_\subC^\subG}{\bar\rho},
\end{align}
where we have applied \eqref{e:J-xeq} to transform into the second line.
Substituting this into \eqref{e:theta-J-2}, we obtain
\begin{align}
&\theta=T_\subC(p)-
\left[J\left(\frac{1}{\kappa_\subC^\subL}-\frac{1}{\kappa_\subC^\subG}\right)+\frac{\phi T_\subC}{L_x}\frac{\rho_\subC^\subL-\rho_\subC^\subG}{\bar\rho}
\right]
\frac{X(L_x-X)}{2L_x}.
\label{e:theta-J-3}
\end{align}
which corresponds to \eqref{e:theta-J} for $J<0$. 
We can repeat the same argument for $J>0$ 
and conclude \eqref{e:theta-J}.

\section{Derivation of  \eqref{e:S*}}
\label{s:additiveS}

We start with an expansion in $\phi$ as
\begin{align}
S(H,p,N,\phi)&=S(H,p,N)+\phi\pderf{S}{\phi}{H,p,N}\!+O(\ep^2)\nonumber\\
&=S(H,p,N)+\phi\Psi+O(\ep^2),
\label{e:S-map}
\end{align}
where we used \eqref{e:th-neq-phi} to obtain the second line.
Let $H_\phi$ be a slightly shifted enthalpy from $H$, whose form
will be determined. Assuming $H-H_\phi=O(\ep)$, we have 
\begin{align}
S(H,p,N)=S(H_\phi,p,N)+\pderf{S}{H}{p,N}\!\!\!(H-H_\phi)+O(\ep^2).
\end{align}
Comparing this form with \eqref{e:S-map}, we find that
the term $\phi\Psi$ in \eqref{e:S-map} cancels out
by setting $H_\phi$ to
\begin{align}
H_\phi=H+T_\subC(p)\phi\Psi.
\label{e:H**}
\end{align}
Thus, we can write 
\begin{align}
S(H,p,N,\phi)=S(H_\phi,p,N)
\label{e:mapping}
\end{align}
in the linear response regime.
We interpret that \eqref{e:mapping} defines an isentropic
surface to connect a heat conduction state $(H,p,N,\phi)$ with
an equilibrium state $(H_\phi,p,N)$.  

Since the right-hand side of \eqref{e:mapping} is the equilibrium entropy, 
the nonequilibrium entropy can be expressed by the sum of the entropy of liquid and gas in \eqref{e:S*}; i.e.,
\begin{align}
S(H,p,N,\phi)
=S(H_\phi^\subL,p,N_\phi^\subL)+S(H_\phi^\subG,p,N_\phi^\subG)
\label{e:S**}
\end{align}
with  $N=N_\phi^\subL+N_\phi^\subG$ and $H_\phi=H_\phi^\subL+H_\phi^\subG$.
We emphasize that $H_\phi \neq H$ indicates $H\neq H_\phi^\subL+H_\phi^\subG$.
This may be another aspect of the violation of the additivity of the liquid--gas coexistence in heat conduction.

We can identify $H_\phi^\subLG$ and $N_\phi^\subLG$ as functions of $(H,p,N,\phi)$. 
Substituting $H_\phi^\subL=\hat h_\subC^\subL(p)N_\phi^\subL$, $H_\phi^\subG=\hat h_\subC^\subG(p)N_\phi^\subG$, and \eqref{e:H**} into 
 $N=N_\phi^\subL+N_\phi^\subG$ and $H_\phi=H_\phi^\subL+H_\phi^\subG$,
 we obtain
 \begin{align}
&N_\phi^\subL  =\frac{\hat h_\subC^\subG(p)N-H -T_\subC(p)\phi\Psi}{\hat q(p)}, \label{e:NL*0}  \\
&N_\phi^\subG=-\frac{\hat h_\subC^\subL(p)N-H-T_\subC(p)\phi\Psi}{\hat q(p)}.\label{e:NG*0}
\end{align}
Comparing \eqref{e:NL*0} with \eqref{e:NL-S3}, we confirm $N_\phi^\subL\neq N^L$ for $\phi\neq 0$.

\vfill



\end{document}